%% file: LMCS-finale-november2013.tex
\documentclass{LMCS}

\def\dOi{9(4:21)2013}
\lmcsheading%
{\dOi}
{1--29}
{}
{}
{Jan.~24, 2013}
{Dec.~12, 2013}
{}

\ACMCCS{[{\bf Mathematics of computing}]: Mathematical
  analysis---Calculus---Lambda calculus; [{\bf Theory of
      computation}]: Models of computation---Computability---Lambda
  calculus}

\usepackage{epsfig}
\usepackage{color}
\usepackage{colortbl}

\usepackage{hyperref}
\usepackage{microtype}
\usepackage{amsmath}
\usepackage{amssymb}
\definecolor{Light}{gray}{.90}

\begin{document}
\input macro-salibra

\title[Ordered models of the lambda calculus]{Ordered models of the lambda calculus\rsuper*}

\author[A.~Carraro]{A.~Carraro\rsuper a}	%required
\address{{\lsuper a}PPS, Universit\'{e} Paris Diderot, B\^{a}timent Sophie Germain 8 Place FM/13 75013 Paris, France}	%required
\email{acarraro@pps.univ-paris-diderot.fr}  %optional
%\thanks{thanks 1, optional.}	%optional

\author[A.~Salibra]{A.~Salibra\rsuper b}	%optional
\address{{\lsuper b}DAIS, Universit\`a Ca'Foscari Venezia, Via Torino 155,  30172 Venezia, Italy}	%optional
\email{salibra@dsi.unive.it}  %optional
\thanks{{\lsuper b}The second author was partially supported by Fondation de Math\'ematique de Paris and 
by the MIUR project ``Metodi logici per il trattamento dell'informazione''}	%optional

%\author[]{Author 3}	%optional
%\address{address 3}	%optional
%\email{author3@email3}  %optional
%\thanks{thanks 3, optional.}	%optional

%% etc.

%% required for running head on odd and even pages, use suitable
%% abbreviations in case of long titles and many authors:

%% mandatory lists of keywords and classifications:
\titlecomment{{\lsuper*}This article is a revised and expanded version of a paper of the authors which appeared  in the
Proceedings of Computer Science Logic (CSL'12) - 21st Annual Conference of the EACSL}
%%%%%%%%%%%%%%%%%%%%%%%%%%%%%%%%%%%%%%%%%%%%%%%%%%%%%%%%%%%%%%%%%%%%%%%%%%%

%% the abstract has to PRECEED the command \maketitle:
%% be sure not to issue the \maketitle command twice!

\begin{abstract}:
\noindent Answering a question by Honsell and Plotkin, we show that there are two equations between $\gl$-terms, the so-called 
 \emph{subtractive equations}, consistent with $\gl$-calculus but not  simultaneously satisfied in any partially ordered model with bottom element.
 We also relate the subtractive equations to the open problem of the order-incompleteness of $\gl$-calculus, by studying the connection
 between the notion of absolute unorderability in a specific point and a weaker notion of subtractivity (namely $n$-subtractivity)
 for partially ordered algebras. Finally we study the relation between $n$-subtractivity and relativized separation conditions in
 topological algebras, obtaining an incompleteness theorem for a general topological semantics of $\lambda$-calculus.
\end{abstract}

\maketitle

\section*{Introduction}\label{S:one}

The lambda calculus was originally introduced by Church \cite{Church33,Church41} as a foundation for logic, where functions, instead of
 sets, were primitive, and it turned out to be consistent and successful as a tool for formalising all computable functions.
 The rise of computers gave a new development to its theoretical studies. The lambda calculus is the kernel of the functional programming
 paradigm, because its ordinary parameter-binding mechanism corresponds closely to parameter binding in many functional programming
 languages.

At the beginning researchers  have investigated lambda calculus by using mainly syntactical methods 
 and have focused their interest on a limited number of equational extensions of lambda calculus, called \emph{$\gl$-theories} (see
 \cite{Bare}). Lambda theories are congruences on the set of $\gl$-terms, which contain $\beta$-conversion. They arise by syntactical or
 semantical considerations. Indeed a $\gl$-theory may correspond to a possible operational semantics of the lambda calculus, as well as it
 may be induced by a model of lambda calculus through the congruence relation associated with the interpretation function. The set of
 $\gl$-theories is naturally equipped with a structure of complete lattice, whose bottom element is the least $\gl$-theory $\gl\gb$, and
 whose top element is the inconsistent $\gl$-theory. The lattice of $\gl$-theories is a very rich and complex structure of cardinality
 $2^{\aleph_0}$ (see, for example, \cite{Bare,LusinS04,ManzonettoS10}). Syntactical techniques are usually difficult to apply in the
 study of $\gl$-theories. Therefore, semantic methods have been extensively investigated.

One of the most important contributions in the area of mathematical programming
semantics was the discovery by Scott \cite{Scott69} in the late 1960s, that complete partial orders, having their own function space as a retract, are models for the untyped lambda calculus. 
It took some time, after Scott gave his model construction, for consensus to arise on the general notion of a model of the $\gl$-calculus. There are mainly two descriptions that one can give: the category-theoretical and the algebraic one. Besides the different languages in which they are formulated, the two approaches are intimately connected (see \cite{Bare}). The categorical notion of model is well-suited for constructing concrete models, while the algebraic one is rather used to understand global properties of models (constructions of new models out of existing ones, closure properties, etc.) and to obtain results about the structure of the lattice of $\gl$-theories.
The algebraic description of models of $\gl$-calculus proposes two kinds of structures, viz. the $\gl$-models based on the notion of combinatory algebra of Curry and Sch\"onfinkel (see  \cite{CurryF58,Schonfinkel24}), and the $\gl$-abstraction algebras of Pigozzi and Salibra (see \cite{PigozziS95,Salibra00,ManzonettoS10}).
Lambda abstraction algebras are intended as an alternative to combinatory algebras since they constitute an equational description of lambda calculus, which keeps the lambda notation and hence all the functional intuitions.

After the construction of the first model by Scott,
a large number of mathematical models for lambda calculus have been introduced in various categories of domains and
were classified into semantics according to the nature of their representable functions \cite{Bare,Berline00,Berline06,Plotkin93}. Scott
continuous semantics \cite{Scott80} is given in the category whose objects are complete partial orders and
morphisms are Scott continuous functions.
Other semantics of lambda calculus were isolated by Berry \cite{Berry78} and Bucciarelli-Ehrhard \cite{BucciarelliE91}:
Berry's stable semantics and Bucciarelli and Ehrhard's strongly stable semantics are refinements of
the continuous semantics introduced to capture the notion of `sequential' Scott continuous function.
All these semantics are structurally and equationally rich \cite{BerlineS06,Kerth95,Kerth01} in the sense that it is possible
to build up $2^{\aleph_0}$-models in each of them, inducing pairwise distinct $\gl$-theories.
On the other hand, there are results that indicate that Scott's methods, based on a combination of order-theory and topology, may not in general be exhaustive.
The problem of the \emph{theory completeness} of a given semantics asks whether
every consistent $\gl$-theory arises as the theory of a model in the semantics.
Honsell and Ronchi Della Rocca \cite{HonsellR92} have shown that there exists a  $\gl$-theory that does not arise as the theory of a Scott model.
Analogous results of incompleteness were obtained by Bastonero and Gouy for the stable semantics \cite{BastoneroG99}  and by Salibra for all models of lambda calculus that involve monotonicity with respect to some pointed (i.e., with bottom element) partial order \cite{Salibra03}.
This last result removes the belief that pointed
partial orderings are intrinsic to models of the lambda calculus, and that the 
incompleteness of a semantics is only due to the richness of the structure of representable functions.
Instead, it is also due to the richness of the structure of the $\gl$-theories.

The need of more abstract semantics of lambda calculus arises when we recognize  the inadequacy of  Scott continuous semantics in order to  investigate the structure of the lattice of $\gl$-theories (see \cite[Chapter 4]{Bare} and \cite{Berline00,Berline06}) in itself and in connections with the theory of models. 
Since topology refines partial orderings through separation axioms,   Salibra \cite{Salibra03} has introduced other topological semantics of lambda calculus,  and has shown: (i) the theory completeness of the semantics of lambda calculus given in terms of metrizable topological models; (ii) the theory incompleteness of the semantics of lambda calculus given in terms of topological models, which have no disjoint closures of nonempty open sets.
This last semantics is enough wide to include properly all pointed ordered models.

A natural problem of \emph{equational completeness} then arises for a semantics of $\gl$-calculus: whether any two $\gl$-terms
equal in all models of the semantics are $\beta$-convertible.  Theory completeness implies equational completeness but the opposite direction is trivially false.  The equational
completeness problem for Scott continuous semantics is one of most outstanding open problems of $\gl$-calculus and it seems to have appeared first in the literature in \cite{HonsellR92}. 
There is also an analogous \emph{equational consistency} problem, raised by Honsell and Plotkin in \cite{HonsellP09}: whether
for every finite set $E$ of equations between $\gl$-terms, consistent with the $\gl$-calculus, there exists a
Scott model contemporaneously satisfying all equations of $E$. In this paper we answer negatively to this second question in a very wide way. We provide two equations (called the \emph{subtractive equations}) consistent with $\gl$-calculus, which  cannot be contemporaneously satisfied by an arbitrary pointed ordered model of $\gl$-calculus. 

Although many familiar models are constructed by order-theoretic methods, it is also known that there are some models of the lambda calculus that cannot be non-trivially ordered (see \cite{Plotkin96,Salibra03,Selinger03}). In general, we define a combinatory algebra $\bA$ to be \emph{unorderable} if there does not exist a non-trivial partial order on $A$ for which the application operation is monotone. Of course, the existence of unorderable models does not imply that order-theoretic methods are somehow incomplete for constructing models: 
an unorderable model can still arise from an order-theoretic construction, for instance as
a subalgebra of some ordered model.
The most interesting result has been obtained by Selinger \cite{Selinger03}, who, 
enough  surprisingly, has shown that the standard open and closed term models of $\gl\gb$ and $\gl\gb\eta$ are unorderable. 
As a consequence of this result, it follows that if $\gl\gb$ or $\gl\gb\eta$ is  the theory of an ordered model, then the denotations of closed terms in that model are pairwise incomparable, i.e. the term denotations form an anti-chain. 
This led Selinger \cite{Selinger03} to study the related question of absolute unorderability: a model is absolutely unorderable if it cannot be embedded in an orderable one. Plotkin conjectures in \cite{Plotkin96} that an absolutely unorderable combinatory algebra exists, but the question is still open whether this is so. Selinger has given in \cite{Selinger03} a syntactic
characterisation of the absolutely unorderable algebras in any algebraic variety (equational class)  in
terms of the existence of a family of Mal'cev operators.  Plotkin's conjecture is thus reduced to the question whether Mal'cev operators are consistent with the lambda calculus or combinatory logic.
 The question of absolute unorderability can also be formulated in terms of theories, rather than models.
In this form, Selinger \cite{Selinger03} refers to it as the \emph{order-incompleteness} question: does there exist a  $\gl$-theory which does not arise as the theory of a non-trivially ordered model?  Such a problem can be also characterised in terms of connected components  of a partial ordering (minimal subsets which are both  upward and downward closed): a  $\gl$-theory $\cT$ is order-incomplete if, and only if, every ordered model, having $\cT$ as equational theory, is partitioned in an infinite number of connected components, each one containing exactly one element. In other words, the partial order is the equality. 

Toward an answer to the order-incompleteness problem, we define  a consistent $\gl$-theory $\cT$ having 
 the following properties: (i) $\cT$ includes  the subtractive equations; (ii)  if an ordered model $\cM$ has an equational theory that contains $\cT$, then $\cM$ has an infinite number of connected components among 
 which that of the looping term $\gO\equiv (\gl x.xx)(\gl x.xx)$ is a singleton set. Moreover, each connected component of $\cM$ contains the denotation of at most one $\beta\eta$-normal form.
Compared to absolute unorderability, the above situation still has some missing bits. For example we are not in the position to tell where the denotations
 of all unsolvable $\gl$-terms other than $\Omega$ are placed in the model. Same thing for all the solvable $\gl$-terms which do not have a $\beta\eta$-normal form.

The inspiration for the subtractive equations comes from the notion of \emph{subtractive variety} of algebras  introduced by Ursini in \cite{Ursini}.  A  variety $\cV$ of algebras is subtractive if it satisfies the following identities:
$$s(x,x) = 0;\qquad s(x,0) = x$$
for some binary term $s$ and constant $0$. Subtractive algebras abound in classical algebras. If we interpret the binary operator ``$s$'' as subtraction, and we use the infix
notation, then we can rewrite the above identities as $x-x= 0$ and $x-0 = x$.
In the context of $\gl$-calculus, the  subtractive equations  make a certain $\gl$-term  behave like a binary subtraction operator (in curryfied form) whose ``zero'' is the looping $\gl$-term $\Omega$. 

We relativize to an element the notion of absolute unorderability.  We say that an algebra $\bA$ is $0$-unorderable if, for every compatible partial order  on $\bA$, $0$ is not comparable with any other element of the algebra. 
An algebra $\bA$ in a variety $\cV$ is  absolutely 0-unorderable
if, for any $\bB\in\cV$ and embedding $f : \bA\to \bB$, $\bB$ is 0-unorderable.
Generalising subtractivity to $n$-subtractivity ($n\geq 2$),
we give a syntactic characterisation of the absolutely $0$-unorderable algebras with Mal'cev-type conditions. The consistency of the two subtractive equations with $\gl$-calculus implies the existence of absolutely $\gO$-unorderable combinatory algebras.

The last result of the paper is a theory incompleteness theorem for topological models of $\gl$-calculus. This result is a strong
 generalisation of an analogous  theorem shown in \cite{Salibra03}. The incompleteness theorem is a consequence of a study of conditions
 of separability for topological $n$-subtractive algebras.

This article is organised as follows: in Section \ref{prelim} we review the basic definitions of universal algebra, topology and $\gl$-calculus
which are involved in the rest of the article. Section \ref{plotkin} is devoted to the proof of consistency of the subtractive equations. In Section \ref{sec:Hon-Plo-question},  we negatively answer to the open question of the equational consistency of Scott continuous semantics. The order incompleteness problem  is presented in Section \ref{subs:strenghtening}. Section \ref{sec:subtractivity} is devoted to the study of the (absolute) $0$-unorderability  in a variety. The main result of Section \ref{sec:subtractivity-topology} is the theory-incompleteness of the semantics of $\gl$-calculus given in terms of topological models, which are not $T_{2_{1/2}}$-separated in $\gO$.

\section{Preliminaries}\label{prelim}

\subsection{Partial Orderings}\label{subsec:poset}

Let $(A,\leq)$ be a partially ordered set (poset). 
Two elements $a,b$ of $A$ are: (1) {\it comparable} 
if either $a \leq b$ or $b\leq a$. 
A set $B\subseteq A$ is an
{\it upward} ({\it downward}) closed set
if $b\in B$, $a\in A$ and $b\leq a$ ($a\leq b$) imply $a\in B$. 

We denote by $\approx_\leq$  the least equivalence relation on $A$ containing $\leq$.
A \emph{connected component} of $(A,\leq)$ is an equivalence class of $\approx_\leq$.
A connected component can be also characterised as a minimal subset of $A$
 which is both upward closed and downward closed.
The poset $(A,\leq)$ is called \emph{connected} if $\approx_\leq$ determines a unique equivalence class. 

\subsection{Algebras}\label{algebras}

An \emph{algebraic similarity type} $\gt$ is constituted by a non-empty set of operator symbols 
together with a function assigning to each operator $f \in\gt$ a finite \emph{arity}.
Operator symbols of arity 0 are called \emph{nullary operators} or \emph{constants}.

 A $\gt$-algebra  $\bA$ is a tuple $(A, f^\bA)_{f\in \gt}$, where $A$ is a non-empty set and 
$f^\bA :A^n\to A$ is an $n$-ary function for every $f\in\gt$ of arity $n$.

Henceforth, the superscript in $f^\bA$ will be dropped whenever the difference
between the operation and the operation symbol is clear from the context, and
a similar policy will be followed in similar cases throughout the paper.

Given two $\gt$-algebras $\bA$ and $\bB$, a {\em homomorphism} from $\bA$ into $\bB$ is a map $g:A\to B$ such that $g(f^{\bA}(a_1,\ldots,a_n)) = f^{\bB}(g(a_1),\ldots,g(a_n))$
for each $n$-ary operator $f\in\gt$ and for all $a_i\in A$.

Given a $\gt$-algebra $\bA$, a binary relation $\phi$ on $A$ is \emph{compatible} if for all $f\in\gt$ of arity $n$,
and for all $a_i,b_i\in A$ we have
$$
    a_1\phi b_1,\ldots, a_n \phi b_n \Longrightarrow f(a_1,\ldots,a_n)\phi f(b_1,\ldots,b_n).
$$
A compatible equivalence relation on an algebra $\bA$ is called a \emph{congruence}.

Let $\cV$ be a class of $\gt$-algebras and $\bA$ be a $\gt$-algebra.

\begin{defi}
If $X\subseteq A$, then we say that $\bA$ has the {\em universal mapping property for $\cV$ over $X$} if, for every $\bB\in \cV$ and for every mapping 
$g: X\to B$, there is a unique homomorphism $f: \bA\to \bB$ that extends $g$ (i.e., $f(x)=g(x)$ for every $x \in X$). 
\end{defi}

\begin{defi}
$\bA$ is {\em free in $\cV$ over $X$} if $\bA\in \cV$, $\bA$ is generated by $X$ under the operations of $\tau$, and $\bA$ has the universal mapping property for 
$\cV$ over $X$. 
\end{defi}

In the following we give a concrete characterization of the
free algebra in the class of all algebras of type $\gt$.

Let $X$ be a set of indeterminates. The set $T_\gt[X]$ of {\em $\gt$-terms}  over $X$ is defined by induction as follows:
\begin{itemize}
\item $x\in T_\gt[X]$ for every $x\in X$;
\item $c\in T_\gt[X]$ for every constant $c\in\gt$;
\item if $t_1,\ldots, t_n\in T_\gt[X]$, then $f(t_1,\ldots,t_n)\in T_\gt[X]$ for all $f\in\gt$ of arity $n$.
\end{itemize}
A $\gt$-term  is \emph{ground} if  it does not contain variables.
If $t$ is a  term, we write $t\equiv t(x_1,\ldots,x_n)$ if the variables occurring in $t$ are among $x_1,\ldots,x_n$.
If $\bA$ is an algebra of type $\gt$ then every  term $t(x_1,\ldots,x_n)$ induces a {\em term operation} $t^\bA:A^n\to A$
defined in the obvious way.

The \emph{free algebra over $X$ in the class of all algebras of type $\gt$}  is the algebra 
$${\bT_\gt}[X] = (T_\gt[X], f^{\bT_\gt[X]})_{f\in \gt},$$
where 
$$f^{\bT_\gt[X]}(t_1,\dots,t_n) = f(t_1,\dots,t_n),\quad \text{for all $f\in \gt$.}$$

An \emph{identity} (or equation) of type $\gt$ is a pair $(t,u)$ of $\gt$-terms, written also $t = u$. A \emph{ground identity} is an equation between ground terms.
An identity $t(x_1,\ldots,x_n) = u(x_1,\ldots,x_n)$ holds in a $\gt$-algebra $\bA$, and we write $\bA\models t = u$,  if the $n$-ary term operations $t^\bA$ and $u^\bA$ are equal.

A non-empty class $\cV$ of algebras of the same type is:
\begin{enumerate}[label=(\roman*)]
\item a \emph{variety} if it is closed under subalgebras, homomorphic images and direct products;
\item  an \emph{equational class} if it is axiomatisable by a set of equations.
\end{enumerate}

\begin{thm} {\rm (Birkhoff \cite{Birkhoff35})} A class of algebras is a variety if, and only if, it is an equational class.
\end{thm}

If $\cV$ is a variety of type $\gt$, then 
 $Eq(\cV)$ is the set of equations satisfied by all algebras in $\cV$.
The set $Eq(\cV)$ is a congruence over the free algebra ${\bf T_\gt}[X]$, and the quotient of $\bT_\gt[X]$ by $Eq(\cV)$ is the free algebra $\bT_\cV[X]$ in $\cV$ over $X$.

\subsection{Topology}\label{top}

By a topological space  we shall mean  a pair $(X,\gt)$, where $X$ is
a non-empty set  and $\gt$ is
a family of subsets of $X$
 closed under infinite unions, finite intersections, and including $\emptyset$ and
$X$.  Given a point $x$ of a space $X$, we say that $U\subseteq X$ is a neighbourhood of $x$ if there exists an open set $V$ such that
$x\in V\subseteq U$.

 A subset $F$ of a space $X$ is closed if $X\setminus F$ is open. The closure of  $U\subseteq X$ will be denoted by $\oU$ (as a matter of notation, we write $\ob$ for
 $\overline{ \{ b\} }$).
Recall that $a\in \oU$ if $U\cap V \neq \emptyset$ for every open
neighbourhood $V$ of $a$.

 A space $X$ is 
\begin{itemize}
\item $T_0$ if, for all distinct $a,b\in X$,  $a$ has a neighbourhood that does not contain $b$ {\bf or} vice versa.
\item $T_1$ if, for all distinct $a,b\in X$, $a$ has a neighbourhood that does not contain $b$ {\bf and} vice versa.
\item $T_2$ (or Hausdorff) if for all distinct $a,b\in X$ there exist open sets
$U$ and $V$ with $a\in U$, $b\in V$ and $U\cap V = \emptyset$.
\item $T_{2_{1/2}}$ (or completely Hausdorff) if for all distinct $a,b\in X$ there exist open sets
$U$ and $V$ with $a\in U$, $b\in V$ and $\oU\cap \oV = \emptyset$.
\end{itemize}

\noindent The previous axioms of separation can be relativized to
pairs of elements.  For example, $a$ and $b$ are
$T_{2_{1/2}}$-separable, if there exist open sets $U$ and $V$ with
$a\in U$, $b\in V$ and $\oU\cap \oV = \emptyset$.  $T_2$-, $T_1$-,
$T_0$-separability are similarly defined.

A preorder can be defined on $(X,\gt)$ by
$$a\leq_\gt b\ \text{iff}\ a\in \ob\  \text{iff}\
\forall U\in \gt (a\in U \Rightarrow b\in U).$$
We have
$$\mbox{$(X,\gt)$ is $T_0$ iff $\leq_\gt$ is a partial order.}$$
For any $T_0$-space $(X,\gt)$ the partial order $\leq_\gt$ is called 
the \emph{specialization order} of $(X,\gt)$. Note that any continuous map between $T_0$-spaces is
necessarily monotone
and that the order is discrete (i.e. satisfies $a\leq_\gt b$ iff $a = b$)
iff $X$ is a $T_1$-space.

 Suppose a space $X$ is not $T_0$. We can obtain a $T_0$-space out of $X$ by the following 
well-known construction. Define an equivalence relation $\equiv$ on $X$ as follows: $a\equiv b\ \Leftrightarrow\ a \leq_\gt b\ \text{and}\ b\leq_\gt a$.
Then $X/\equiv$ equipped with the quotient topology is a $T_0$-space.

A space $(X,\gt)$ is \emph{coconnected} if for all non-empty opens $U, V\in \gt$ we have $\oU\cap\oV\neq\emptyset$ (see \cite{Salibra03}).

\subsection{Topological algebras}

A {\it topological algebra}  is a pair $(\bA,\gt)$
where $\sbA$ is an algebra and $\gt$ is a topology on the underlying set $A$
with the property that each basic operation $f^\bA: A^n\to A$ of $\bA$ is continuous,
where the domain of $f$ is endowed with the product topology.

A {\it semitopological algebra}  is a pair $(\bA,\gt)$
where $\sbA$ is an algebra and $\gt$ is a topology on the underlying set $A$
with the property that each unary polynomial of $\sbA$ is
continuous w.r.t. $\gt$.
Every topological algebra is a semitopological algebra.

A map $f: A^n \to A$ is \emph{separated-continuous} if, for every $1 \leq i \leq n$
and every $a_1,\dots,a_{i-1},a_{i+1},\dots,a_n \in A$,
the map $g:A\to A$, defined by
$$g(b) = f(a_1,\dots,a_{i-1},b,a_{i+1},\dots,a_n),\quad\mbox{for all $b\in A$},$$
is continuous.
In a semitopological algebra every term operation is separated-continuous.

\begin{lem} \label{1.3.1bis}
Let $(A,\gt)$ be a topological space. Then every (separated-)continuous map $f: A^n \to A$ is monotone:
$$a_i\leq_\gt b_i\ (1\leq i\leq n)\ \Rightarrow\ f(a_1,\dots,a_n) \leq_\gt f(b_1,\dots,b_n).$$
\end{lem}

\proof
  For the sake of simplicity, assume $f$ is binary.
Let $a \leq_\gt b$ and $a' \leq_\gt b'$. We have to show that $f(a,a') \leq_\gt f(b,b')$.
Let $f(a,a')\in U$, where $U$ is open. Then by continuity in first coordinate there exists
an open set $V$ such that $a\in V$ and $f(V,a') \subseteq U$. By $a\leq_\gt b$ it follows that
$b\in V$ so that $f(b,a')\in U$. By using continuity in second coordinate we get that there
exists an open set $W$ such that $a'\in W$ and $f(b,W)\subseteq U$. By $a'\leq_\gt b'$ we get $b'\in W$
and then the conclusion.
\qed

If $(\sbA,\gt)$ is a semitopological algebra then $\leq_\gt$ is a compatible preorder.

%%%%%%%%%%%%%%%%%%%%%%%%%%%%%%%%%%%%%%%%%%%%%%%%%%%%%
\subsection{Lambda calculus}\label{sus:lam-calc}
%%%%%%%%%%%%%%%%%%%%%%%%%%%%%%%%%%%%%%%%%%%%%%%%%%

With regard to the $\gl$-calculus we follow the notation and terminology of \cite{Bare}. By $\Lambda$ and $\Lambda^o$, respectively, we indicate
 the set of $\gl$-terms and of closed $\gl$-terms. By convention application associates to the left.
The symbol $\equiv$ denotes syntactical equality. The following are some notable $\gl$-terms that will be used throughout the paper: $\gO \equiv (\gl x.xx)(\gl x.xx)$; $\ssi \equiv \gl x.x$; $\ssT \equiv \gl xy.x$; $\ssF \equiv \gl xy.y$.

If $M$ is a $\gl$-term and $\vec{P}\equiv P_1\dots P_n$ is a sequence of $\gl$-terms, we write $M\vec P$ for the application $MP_1\cdots P_n$.
 
The $\beta$-reduction will be denoted by $\labelra{\beta}$, while the $\eta$-reduction by $\labelra{\eta}$.
 One step of either $\beta$-reduction or $\eta$-reduction will be denoted by $\labelra{\beta\eta}$. 
 
 A $\gl$-term $M$ is \emph{solvable} if it has a \emph{head normal form}, i.e., $M$ is $\beta$-convertible to a term of the
 form $\lambda \vec{x}.y\vec{N}$. A $\gl$-term $M$ is \emph{unsolvable} if it is not solvable.

 A {\em $\gl$-theory} is a congruence on $\Lambda$ (with respect to the operators of abstraction and application) which contains $\ga\gb$-conversion.
 We denote  by $\gl\gb$ the least $\gl$-theory. The least extensional $\gl$-theory  $\gl\gb\eta$ is axiomatised over $\gl\gb$ by the
 equation $\gl x. Mx = M$, where $M\in \gL$ and $x$ is not free in $M$. 
 The $\gl$-theory generated by a set $X$ of identities is the intersection of all $\gl$-theories containing $X$.
 For a $\gl$-theory $\cT$ we will write $\cT\vdash M=N$ (or $M=_\cT N$) to indicate the existence of an equational proof of the identity
 $M=N$ that uses equations of $\cT$.

 A $\gl$-theory is {\em consistent} if it does not equate all $\gl$-terms, {\em inconsistent} otherwise. It turns out that a $\gl$-theory
 $\cT$ is inconsistent iff $\cT\vdash \ssT = \ssF$. A $\gl$-theory is \emph{semisensible} if it does not equate solvable and unsolvable
 $\gl$-terms. Semisensible $\gl$-theories are consistent (see \cite{Bare}). The set of $\gl$-theories constitutes a complete lattice
 w.r.t. inclusion, whose top is the inconsistent $\gl$-theory and whose bottom is the theory $\lambda\beta$.

 An unsolvable is called a \emph{zero term} if it never reduces to an abstraction. 
 A zero term, that will be used in the rest of the paper, is defined as
 follows. Let $B \equiv \gl x.x(\gl y.yx)$, $C \equiv \gl z.zB$ and $\Theta \equiv BC$. By a direct computation we see that the only 
 possible reduction path starting with $\Theta$ is the following:
$$\Theta \labelra{\beta} C(\gl y.yC) \labelra{\beta} (\gl y.yC)B \labelra{\beta} BC \equiv \Theta.$$

 The letters $\gx_1,\gx_2,\dots$ denote algebraic variables (holes in Barendregt's terminology \cite{Bare}). Contexts are built up as $\gl$-terms but also allowing occurrences of algebraic variables. Substitution for algebraic variables is made without $\ga$-conversion. For example,
$(\gl x. x\gx)[xy/\gx] = \gl x.x(xy)$.

 We recall the Genericity Lemma of lambda calculus  (see \cite[Proposition 14.3.24]{Bare}).

\begin{lem}\label{gen}
 Let $M\in \gL$ unsolvable and $N\in \gL$ having a normal form. Then, for every context $C[\gx]$, 
 $$C[M] =_{\gl\gb} N\ \Longrightarrow\ (\forall Q\in \gL)\ C[Q] =_{\gl\gb} N.$$  
\end{lem}

Throughout the paper we consider different reductions. If $\labelra{\gamma}$ is a reduction, then we denote by $\mslabelra{\gamma}$ the
 reflexive transitive closure of $\labelra{\gamma}$, and we write $=_\gamma$ to denote the reflexive, symmetric and transitive closure of
 $\labelra{\gamma}$. Finally we define, as usual, the \emph{$\gamma$-reduction graph} of a term $M$ as the set
 $\cG_\gamma(M) = \{N \in \Lambda : M \mslabelra{\gamma} N\}$.

%%%%%%%%%%%%%%%%%%%%%%%%%%%%%%%%%%%%%%%%%%%%%%%%%%%%%%%%%%%%%%%%%%%%%%%%%%%%%%%%%%%%%%%%%%%%%%%%%%%%%%%%%%%%%%%%%%%%%%%%%%%%%%%%%%%%%%
\subsection{Models of $\lambda$-calculus}
%%%%%%%%%%%%%%%%%%%%%%%%%%%%%%%%%%%%%%%%%%%%%%%%%%%%%%%%%%%%%%%%%%%%%%%%%%%%%%%%%%%%%%%%%%%%%%%%%%%%%%%%%%%%%%%%%%%%%%%%%%%%%%%%%%%%%%%

The algebraic description of models of $\lambda$-calculus proposes two kinds of structures, viz. the \emph{$\gl$-algebras} and the
 \emph{$\gl$-models}, both based on the notion of \emph{combinatory algebra}. We will focus on $\gl$-models.
A \emph{combinatory algebra} $\bA = (A,\cdot,K, S)$ is a structure with a binary operation called \emph{application} and two distinguished
 elements $K$ and $S$ called \emph{basic combinators}. The symbol ``$\cdot$'' is usually omitted from expressions and by convention
 application associates to the left, allowing to leave out superfluous parentheses. The class of combinatory algebras is axiomatised by
 the equations $K xy = x$ and $S xyz = xz(yz)$. Intuitively, elements on the left-hand side of an application are to be seen as functions
 operating on arguments, placed on the right-hand side. Hence it is natural to say that a function $f:A^n \to A$ is \emph{representable}
 (\emph{in} $\bA$) if there exists an element $a \in A$ such that $f(b_1,\dots,b_n) = ab_1\dots b_n$ for all $b_1,\dots, b_n \in A$. For example the identity function is
 represented by the combinator $I\equiv SKK$ and the  projection on the first argument by the combinator $K$.

The axioms of an elementary subclass of combinatory algebras, called \emph{$\lambda$-models}, were expressly chosen to make coherent the definition of interpretation of $\gl$-terms. In addition to the axioms of combinatory algebra, we have: 
 \[
\begin{array}{lll}
 &  \forall xy.(\forall z.\ xz = yz) \Rightarrow \sso x = \sso y & \\
 & \sso_2K = K &   \\
 & \sso_3S = S &
\end{array}
\]
where $\sso_1 \equiv \sso \equiv S(KI)$ and $\sso_{n+1} \equiv S(K\sso)(S(K\sso_n))$.
 The combinators $\sso_n$ are made into  inner choice operators. Indeed, given any $a \in A$, the element $\sso_n a$ represents the same $n$-ary
 function as $a$ and $\sso_n c = \sso_n d$ for all $c,d$ representing the same $n$-ary function.

Let $Env_A$ be the set of $A$-environments, i.e.\ , the functions from the set $\mathrm{Var}$ of $\gl$-calculus variables to $A$.
 For every $x\in \mathrm{Var}$ and $a\in A$ we denote by $\gr[x:=a]$ the environment $\gr'$ which coincides with $\gr$ everywhere except on $x$,
 where $\gr'$ takes the value $a$. 
 
When $\bA$ is a $\lambda$-model it is possible to define the following interpretation:
\[
\begin{array}{ll}
{|x|}_\gr^\bA  = \gr(x); &\\
{|MN|}_\gr^\bA = {|M|}_\gr^\bA {|N|}_\gr^\bA; &\\
{|\gl x.M|}_\gr^\bA = \sso a,\ \text{where $a\in A$ is any element representing the function $b\in A \mapsto {|M|}_{\gr[x:=b]}^\bA$.}&
\end{array}
\]
Note that ${|\gl x.M|}_\gr^\bA$ is well-defined, since each function $b\in A \mapsto {|M|}_{\gr[x:=b]}^\bA$ is representable under the hypothesis that $\bA$ is a $\gl$-model. This is the kind of interpretation we will refer to. 

By the way when $M$ is a closed $\gl$-term and there is no worry of confusion about the model being considered, we write $|M|$ for ${|M|}_\gr^\bA$.

Each $\lambda$-model $\bA$ induces a $\gl$-theory, denoted here by $Th(\bA)$, and called {\em the equational theory of $\bA$}.
 Thus, $M = N \in Th(\bA)$ if, and only if, $M$ and $N$ have the same interpretation in $\bA$.

An \emph{ordered $\gl$-model} is a pair $(\bA,\leq)$, where $\bA$ is a $\gl$-model and $\leq$ is a partial order on $A$ which makes the application operator of $\bA$ monotone in both arguments. 
An ordered model $(\bA,\leq)$ is \emph{non-trivial} if the partial order
is not discrete, i.e., $a<b$ for some $a,b\in A$ (thus $A$ is not a singleton).
A \emph{pointed ordered model} is an ordered model with bottom element.

The {\it term model} $\cM_\cT$ of a lambda theory $\cT$ (see \cite[Def.~5.2.11]{Bare})
 consists of the set of the equivalence classes of $\gl$-terms modulo the lambda theory $\cT$ 
 together with the operation of application on the equivalence classes.
 By Cor.~5.2.13(ii) in \cite{Bare} $\cM_\cT$ is a $\gl$-model such that $Th(\cM_\cT)= \cT$.

A class $\Cc$ of $\gl$-models is 
\begin{enumerate}%[(1)]
\item \emph{equationally consistent} if, for every finite set $E$ of equations between $\gl$-terms, consistent with the $\gl$-calculus, there exists a
 model  $\bA\in \Cc$ simultaneously satisfying all equations of $E$.
 \item \emph{equationally complete} if any two $\gl$-terms
equal in all models of $\Cc$ are $\beta$-convertible.
 \item \emph{theory complete} if, for every consistent $\gl$-theory $\cT$, there exists a model $\bA\in \Cc$ such that $Th(\bA) = \cT$.
 \end{enumerate} 

%%%%%%%%%%%%%%%%%%%%%%%%%%%%%%%%%%%%%%%%%%%%%%%%%%%%%%%%%%%%%%%%%%%%%%%%%%%%%%%%%%%%%%%%%%%%%%%%%%%%%%%%%%%%%%%%%%%%%%%%%%%%%%%%%%%%%%%
\subsection{The Jacopini\textendash Kuper technique}\label{subs:Jacopini-Kuper}
%%%%%%%%%%%%%%%%%%%%%%%%%%%%%%%%%%%%%%%%%%%%%%%%%%%%%%%%%%%%%%%%%%%%%%%%%%%%%%%%%%%%%%%%%%%%%%%%%%%%%%%%%%%%%%%%%%%%%%%%%%%%%%%%%%%%%%%

The Jacopini\textendash Kuper technique, introduced by Jacopini in \cite{Ja} and generalized by Kuper in \cite{Kuper97}, can be used to tackle questions of consistency of equational extensions of lambda calculus.
 In this section we review this technique.

Let $\cT$ be an arbitrary consistent $\lambda$-theory, $\vec P = \vec Q$ be a set of identities $P_i = Q_i$ ($i=1,\dots, n$) between closed $\gl$-terms, and  $\cT'$ be the  $\gl$-theory generated by $\cT\cup (\vec P = \vec Q)$.
The idea is to reduce inconsistency of $\cT'$ to that of $\cT$. If $\cT'$ is inconsistent, then there exists a finite equational proof of $\ssT =_{\cT'} \ssF$,
 and such a proof contains a finite number of applications of equations which are in $\vec P = \vec Q$. 
 Jacopini\textendash Kuper technique, when applicable, consists in checking two conditions on the sequences $\vec P$ and $\vec Q$, namely 
 that $\vec P$ is $\cT$-\emph{operationally less defined} than $\vec Q$ (see Definition \ref{def:op-def}) and
 that $\vec P$ is $\cT$-\emph{proof-substitutable} by $\vec Q$ (see Definition \ref{def:proof-subst}).
 Under these two conditions, it is possible to remove from the proof of $\ssT =_{\cT'} \ssF$ all occurrences of 
 equations  in $\vec P = \vec Q$, thus yielding a proof of $\ssT =_{\cT} \ssF$. This is the end of the method, because $\cT$ is supposed to be a consistent $\lambda$-theory.
 
A very useful property for the application of Jacopini-Kuper method, and in particular for proving
 $\cT$-proof-substitutability, is the existence of a Church-Rosser reduction, 
  whose induced conversion coincides with the equality induced by $\cT$ on $\lambda$-terms.
 This is not evident from the abstract formulation given in this section, but will be clear in the next one, when we will apply the technique.

\begin{lem}\label{lem:standard-eq-proof}
We have that 
$\cT\cup (\vec P = \vec Q) \vdash M=N$
if, and only if, there exist closed terms $F_1,\dots, F_n$ such that
 \[
 \begin{array}{llll}
M                 & =_{\cT} & F_1\vec P\vec Q& \\ 
F_{j}\vec Q\vec P & =_{\cT} & F_{j+1}\vec P\vec Q &\text{for $1\leq j\leq n-1$}\\
F_n\vec Q\vec P   & =_{\cT} & N.&
\end{array}
\]
\end{lem}

\proof By \cite[Theorem 1]{Du} 
 there exist binary contexts $C_1(\gx_1,\gx_2), \dots, C_n(\gx_1,\gx_2)$  and identities $P_{i_j} = Q_{i_j}$ in $\vec P = \vec Q$ such that 
 \[
\begin{array}{llll}
M  & =_\cT  & C_1(P_{i_1},Q_{i_1}) & \\
C_j(Q_{i_j}, P_{i_j} )  & =_\cT  &C_{j+1}(P_{i_{j+1}},Q_{i_{j+1}}) &\text{for $1\leq j\leq n-1$} \\
C_n(Q_{i_n}, P_{i_n})  &  =_\cT & N.  &
\end{array}
\]
It is sufficient to define $F_j \equiv \gl \vec x\vec y. C_j(x_{i_j},y_{i_j})$,
where $\vec x$ and $\vec y$ are sequences of length $k$ of fresh variables.
\qed

\begin{defi}[Operational definiteness]\label{def:op-def}
 We say that $\vec P$ is $\cT$-operationally less defined than $\vec Q$
if, for every $\gb\eta$-normal form $N$ and every term $F$, we have that
 $$ F\vec P=_\cT N \Rightarrow F\vec Q =_\cT N. $$
\end{defi}

\begin{defi}[Proof-substitutability]\label{def:proof-subst}  We say that $\vec P$ is $\cT$-proof-substitutable by $\vec Q$ if
$$\forall F,F'\in\Lambda^o(F \vec P =_{\cT} F'\vec P \Rightarrow
  \exists G   \in\Lambda^o(G \vec P\vec Q =_{\cT} F \vec Q \mbox{ and }
                           G \vec Q\vec P =_{\cT} F'\vec Q)).$$
\end{defi}

\begin{thm}\label{thm:reduce-incons}
If $\vec P$ is $\cT$-operationally  less defined than $\vec Q$ and $\vec P$ is $\cT$-proof-substitutable by $\vec Q$, then the $\gl$-theory $\cT'$ generated by $\cT\cup (\vec P = \vec Q)$ is consistent.
\end{thm}

\proof
 Assume $\cT'$ is inconsistent, so that $\ssT =_{\cT'} \ssF$.
 Then by Lemma \ref{lem:standard-eq-proof} there exists an equational proof of this identity of the form
\[
\begin{array}{llll}
\ssT            & =_{\cT} & F_1\vec P\vec Q& \\ 
F_{j}\vec Q\vec P    & =_{\cT} & F_{j+1}\vec P\vec Q& \text{for $1\leq j\leq n-1$}\\
 F_n\vec Q\vec P      & =_{\cT} & \ssF.&
\end{array}
\]
Now we show how to iteratively transform the above proof of $\cT'\vdash \ssT =\ssF$ in a proof of $\cT\vdash \ssT=\ssF$.

Suppose $n = 1$, i.e., we have
$\ssT   =_{\cT} F_1\vec P\vec Q$ and
$F_{1}\vec Q\vec P =_{\cT} \ssF$.
Since $\vec P$ is $\cT$-operationally less defined than $\vec Q$, from
 $\ssT =_{\cT} F_1\vec P\vec Q =_{\cT} (\gl \oy. F_1\oy \vec Q) \vec P $ and $F_{1}\vec Q\vec P =_{\cT} \ssF$, we get 
 $\ssT =_{\cT} F_1\vec Q\vec Q =_{\cT}  \ssF$.

Suppose $n > 1$. As before, by the hypothesis we get $\ssT =_{\cT} F_1\vec Q\vec Q$ 
 and $F_n\vec Q\vec Q =_{\cT} \ssF$. 
Let $\vec y$ be a sequence of fresh variables.  For each $j = 1,\ldots,n-1$, define
$$H_{j} \equiv \lambda \vec y.F_{j}\vec Q\vec y;\qquad H_{j+1}' \equiv \lambda \vec y.F_{j+1}\vec y\vec Q.$$
By  $F_{j}\vec Q\vec P =_{\cT} F_{j+1}\vec P\vec Q$ ($j = 1,\ldots,n-1$) we have that
 $$H_{j}\vec P =_{\cT} H_{j+1}'\vec P.$$
Since $\vec P$  is  $\cT$-proof-substitutable by $\vec Q$, then   
 there exist terms $G_{j}$ ($j = 1,\ldots,n-1$) such that 
 $G_{j}\vec P\vec Q =_{\cT} H_{j}\vec Q =_{\cT} F_{j}\vec Q\vec Q$ and
 $G_{j}\vec Q\vec P =_{\cT} H_{j+1}'\vec Q =_{\cT} F_{j+1}\vec Q\vec Q$.
Therefore we obtain that
\[
\begin{array}{rclcl}
\ssT           & =_{\cT} & F_1\vec Q\vec Q   & =_{\cT} & G_1\vec P\vec Q \\
G_{1}\vec Q\vec P   & =_{\cT} & F_{2}\vec Q\vec Q & =_{\cT} & G_2\vec P\vec Q \\
                    &     \vdots   &                   & \vdots       & \\
G_{n-1}\vec Q\vec P & =_{\cT} & F_{n}\vec Q\vec Q & =_{\cT} & \ssF \\
\end{array}
\]
Therefore one can iterate the argument and get a proof of $\cT\vdash \ssT=\ssF$.
\qed

%%%%%%%%%%%%%%%%%%%%%%%%%%%%%%%%%%%%%%%%%%%%%%%
\section{On a question by Honsell and Plotkin}\label{plotkin}
%%%%%%%%%%%%%%%%%%%%%%%%%%%%%%%%%%%%%

In this section we turn to a question posed in \cite{HonsellP09} by Honsell and Plotkin. The problem is whether or not there exists 
 a formula $\varphi$ of first-order logic written as a possibly empty list of universal quantifiers followed by a conjunction of
 equalities between $\lambda$-terms such that $\varphi$ does not admit pointed ordered models. According to Honsell and Plotkin, this problem falls
 under the name of $\Pi_1$-consistency of the class of pointed ordered models. 
  Since any equation between $\gl$-terms is equivalent to a suitable equation between closed $\gl$-terms, 
 then in the context of $\lambda$-calculus the $\Pi_1$-consistency is equivalent to the equational consistency, 
 which is the particular case of $\Pi_1$-consistency in which the formula $\varphi$ is quantifier-free.  

%%%%%%%%%%%%%%%%%%%%%%%%%%%%%%%%%%%%%%%%%%%%%%%%%%%%%
\subsection{The $\gl$-theory $\lambda\pi\phi$}\label{sec:theory-lpp}
%%%%%%%%%%%%%%%%%%%%%%%%%%%%%%%%%%%%%%%%%%%%%%%%%%%%%

We introduce two equations between $\gl$-terms, whose models will be shown to have strong properties with respect to the possible partial orderings they can be endowed with. Of course we  have to prove that the $\gl$-theory $\lambda\pi\phi$ generated by these equations is consistent
 and this will be done in Section \ref{subs:Jac-for-lpp}.

The two equations we are going to introduce represent within $\gl$-calculus the notion of subtractivity, which has been introduced in Universal Algebra 
by Ursini \cite{Ursini}.

\begin{defi}\label{sub} An algebra $\bA$ is \emph{subtractive} if there exist a binary term
$s(x,y)$ and a constant $0$ in the algebraic similarity type of $\bA$ such that
$$s(x,x) = 0;\qquad s(x,0)=x.$$
\end{defi}

Subtractive algebras abound in classical
algebras and in algebraic logic since the term $s$ simulates part of subtraction. If
we interpret the binary operator ``$s$'' as subtraction ``$-$'' and we use the infix
notation, then we can rewrite the above identities as $x-x= 0$ and $x-0 = x$.

The following lemma explains why we are interested in subtractive equations. 

 \begin{lem}\label{lem-ord} Let $\bA$ be a subtractive algebra and $\leq$ be a compatible partial order on $A$. Then,  every element $a\neq 0$ of $A$ is incomparable with   $0$.
\end{lem}

\proof If $a\leq 0$ then $0= s(a,a) \leq s(a,0)= a$. 
If $0\leq a$ then $a = s(a,0) \leq s(a,a) = 0$.
\qed

A general treatment of subtractivity and orderings in Universal Algebra can be found in Section \ref{sec:subtractivity}.   

We now define the $\gl$-theory $\lambda\pi\phi$. 
Let $\Theta$  be the  term defined in Section \ref{sus:lam-calc}.  We define $s(x,y) \equiv \Theta xy$ and $0\equiv \gO$. 
Then the  $\gl$-theory  $\lambda\pi\phi$ is defined as the least extensional $\lambda$-theory generated by the following two equations, called the \emph{subtractive equations}:
$$(\pi)\  \  \Theta xx = \gO;\quad\qquad (\phi)\ \ \Theta x \gO = x.$$
The intuitive meaning of the equations ($\pi$) and ($\phi$) is that they make the term $\Theta$ behave like a 
 binary subtraction operator (in curried form) whose ``zero'' is the term $\Omega$. 
 We have chosen $\Theta$ to represent the binary subtraction operator because the reduction graph of $\Theta$ is as simple as possible among the unsolvables distinct from $\gO$.

 The following theorem illustrates a curious aspect of the equations ($\pi$) and ($\phi$):  the choice of $\gO$ is the right one.

\begin{thm}\label{thm:must-be-Omega} Let $O$ be a $\gl$-term such that $x\notin FV(O)$, and let $\cT$ be any $\gl$-theory including the identities $\Theta x O = x$ and $\Theta xx = O$. Then $\cT \vdash O =Y\ssi$ for every  fixpoint combinator $Y$. In particular,  if $Y\equiv \lambda f.(\lambda x.f(xx))(\lambda x.f(xx))$ is the Curry fixpoint combinator, then $\cT \vdash O = \Omega$.
\end{thm}

\proof 
We apply a technique introduced by Gordon Plotkin and Alex Simpson (see \cite{Selinger03}).
Let $Y$  be an arbitrary  fixpoint combinator.
 Then, for any $\lambda$-term $M$, define $\mu x.M \equiv Y(\lambda x.M)$. 
Now let $D \equiv \mu y.\mu x. \Theta xy$. Then we have $D =_\beta \Theta DD =_\cT O$ and
 $D=_\beta \mu x. \Theta xD =_\cT\mu x. \Theta xO =_\cT \mu x.x \equiv Y\ssi$. Moreover, $Y\ssi =_\gb \gO$ if $Y$ is the Curry fixpoint combinator.
\qed

%%%%%%%%%%%%%%%%%%%%%%%%%%%%%%%%%%%%%%%%%%%%%%%%%%
\subsection{The $\gl$-theory $\lambda\pi$}\label{subs:lp}
%%%%%%%%%%%%%%%%%%%%%%%%%%%%%%%%%%%%%%%%%%%%%%%

The extensional $\lambda$-theory $\lambda\pi$ is axiomatised over $\gl\gb\eta$  by the equation $(\pi)$. It is consistent because semisensible.
We will show the consistency of $\lambda\pi\phi$ relying on the consistency of $\lambda\pi$. 

The following notion of reduction will be useful in the next section to prove the consistency of the $\gl$-theory $\lambda\pi\phi$
(recall from  Section \ref{sus:lam-calc} the definition of $\Theta$ and of its reduction graph $\cG_\beta(\Theta)$).

\begin{defi}[$\lambda\pi$-reduction]
We introduce here $\beta\eta\pi$-reduction, notation $\labelra{\beta\eta\pi}$, as the contextual closure of $\labelra{\beta\eta} \cup \labelra{\pi}$, where
$$\Psi M N \labelra{\pi} \Omega\quad\text{if $\Psi\in \cG_\beta(\Theta)$ and  $\lambda\pi \vdash M = N$}.$$
\end{defi}

Of course the conversion $=_{\beta\eta\pi}$ coincides with the equality induced by ${\lambda\pi}$.
 We remark that in \cite{Salibra03} it was introduced a $\gl$-theory axiomatised by $\gO xx = \gO$.
 Here we use instead $\Theta xx = \gO$ for technical reasons. In fact when we apply the Jacopini\textendash Kuper technique in
 Section \ref{subs:Jac-for-lpp}, we use the fact that a step of reduction $\Theta x x \labelra{\pi} \gO$ does not create a new
 $\labelra{\pi}$-redex, because $\gO \not\in \cG_\beta(\Theta)$.

Note that if a $\gl$-term $N$ is a $\beta\eta$-normal form, then it is also a $\beta\eta\pi$-normal form, because every
 $\labelra{\pi}$-redex contains also a $\labelra{\beta}$-redex.

\begin{thm}\label{thm:betapil-confluent}\hfill
\begin{enumerate}[label=(\roman*)]
\item The reduction $\labelra{\beta\eta\pi}$ is Church-Rosser;
\item For all terms $M$ and $N$, we have $\Theta MN =_{\beta\eta\pi} \Omega$ iff $\lambda\pi \vdash M = N$.
\end{enumerate}
\end{thm}

\proof\hfill
\begin{enumerate}[label=(\roman*)]
\item First observe that the relation $\mslabelra{\pi}$ is Church-Rosser.
 Moreover the relations $\mslabelra{\pi}$ and $\mslabelra{\beta\eta}$ commute, meaning that 
 whenever $Q \mslabella{\pi} P \mslabelra{\beta\eta} Z$ there exists $P'$ such that $Q \mslabelra{\beta\eta} P' \mslabella{\pi} Z$.
 The conclusion follows from the Hindley\textendash Rosen Lemma (see \cite[Prop.~3.3.5]{Bare}), which states that
 if two Church-Rosser relations commute, then their union is Church-Rosser too.

\item If $\Theta MN =_{\beta\eta\pi} \Omega$ then by (i) there exists a reduction $\Theta MN \mslabelra{\beta\eta\pi} \Omega$.
 But this is possible only if $\Theta MN \mslabelra{\beta\eta\pi} \Psi M'N'$ with $\Theta \mslabelra{\beta} \Psi$,
 $M \mslabelra{\beta\eta\pi} M'$, $N \mslabelra{\beta\eta\pi} N'$ and $\lambda\pi \vdash M'= N'$. Therefore 
 $\lambda\pi \vdash M=N$. \qed
\end{enumerate}

\noindent Another useful result is the forthcoming lemma, which says that all ${\beta\eta\pi}$-reduction paths may be ``simulated''
 by a reduction path which contains ${\pi}$-steps only at the end. We indicate by $\to^=_{\beta\eta}$ the reflexive closure of
 $\labelra{\beta\eta}$. 

\begin{lem}[Factorization]\label{lem:factorization}
If $M \mslabelra{\beta\eta\pi} N$, then there exists $P$ such that $M \mslabelra{\beta\eta} P \mslabelra{\pi} N$.
\end{lem}

\proof
Use iteratively the fact that whenever $M \labelra{\pi} N \labelra{\beta\eta} Q$, then there exists $N'$
such that  $M \to^=_{\beta\eta} N' \mslabelra{\pi} Q$.
\qed

\begin{lem}\label{lem:theta-not-omega}
The terms $\Theta$ and $\Omega$ are not ${\beta\eta\pi}$-convertible.
\end{lem}

\proof
By the reduction graph of $\Theta$ and the confluence of $\labelra{\beta\eta\pi}$.
\qed

%%%%%%%%%%%%%%%%%%%%%%%%%%%%%%%%%%%%%%%%%%%%%%%%%%%%%%%%%%%%%%%%%%%%%%%%%%%%%%%%%
\subsection{Jacopini\textendash Kuper technique for $\lambda\pi\phi$}\label{subs:Jac-for-lpp}
%%%%%%%%%%%%%%%%%%%%%%%%%%%%%%%%%%%%%%%%%%%%%%%%%%%%%

In this section we apply the Jacopini\textendash Kuper technique explained in Section \ref{subs:Jacopini-Kuper} to prove the consistency 
 of the theory $\lambda\pi\phi$. 
 More precisely, the results presented here show that the closure $\gl x.\Theta x\gO = \ssi$ of the equation $(\phi)$,
 that axiomatizes $\lambda\pi\phi$ over $\lambda\pi$, satisfies the hypotheses of Theorem \ref{thm:reduce-incons}.

\begin{lem}\label{lem:bottom-preorder}
The term $\lambda x.\Theta x \Omega$ is $\lambda\pi$-operationally less defined than $\ssi$.
\end{lem}

\proof
Let $F$  be a $\lambda$-term and $N$ be a $\gb\eta$-normal form, and suppose
 \mbox{$F(\lambda x.\Theta x \Omega) =_{\lambda\pi} N$}.
 Since ${\lambda\pi}$-reduction is confluent and $N$ is $\gb\eta$-normal, we have that
 \mbox{$F(\lambda x.\Theta x\Omega) \mslabelra{\beta\eta\pi} N$}.
 By Lemma \ref{lem:factorization} there exists a term $M$ such that
$$F(\lambda x.\Theta x \Omega) \mslabelra{\beta\eta} M \mslabelra{\pi} N.$$  
Since $N$ is a $\gb\eta$-normal form and the reduct $\gO$ of a $\labelra{\pi}$-redex is a $\gb$-redex,
 we must have that $M \equiv N$. Therefore we have $\lambda\beta\eta \vdash F(\lambda x.\Theta x \Omega) = N$ and, since 
 $\lambda x.\Theta x \Omega$ is unsolvable, we can apply the Genericity Lemma of lambda calculus  (see Lemma \ref{gen}) to obtain 
 $\lambda\beta\eta \vdash F\ssi = N$, and hence obviously $\lambda\pi \vdash F\ssi = N$ which is the desired conclusion.
\qed

In Lemma \ref{lem:proof-subst} below we keep track of the  residuals of the  $\lambda$-term $\lambda x.\Theta x \Omega$ during the reduction of  the term
 $F(\lambda x.\Theta x \Omega)$. We have three kinds of residuals: $\gl x.\Psi x\gO$, $\Psi M\gO$ and $\gO$ (with $\Psi\in \cG_\beta(\Theta)$) as the following informal example shows:
 \[
\begin{array}{llll}
F(\lambda x.\Theta x \Omega) &\mslabelra{\gl\pi} &\cdots (\lambda x.\Theta x \Omega)\cdots (\lambda x.\Theta x \Omega)\cdots (\lambda x.\Theta x \Omega)\cdots  &\\
           &  \mslabelra{\gb}   &\cdots (\lambda x.\Theta x \Omega)\cdots (\lambda x.\Psi x \Omega)\cdots\cdots (\lambda x.\Theta x \Omega)\cdots&\text{($\Theta\mslabelra\gb \Psi$)} \\
           &  \mslabelra{\gl\pi}   &\cdots (\lambda x.\Theta x \Omega)\cdots (\lambda x.\Psi x \Omega)M\cdots (\lambda x.\Theta x \Omega)\cdots& \\
           &  \labelra{\gb}   &\cdots (\lambda x.\Theta x \Omega)\cdots (\Psi M \Omega)\cdots (\lambda x.\Theta x \Omega)\cdots &\text{($\gb$-reduction)} \\
           &  \mslabelra{\gl\pi}   &\cdots (\lambda x.\Theta x \Omega)\cdots (\Psi N \Omega)\cdots  (\lambda x.\Theta x \Omega)\cdots&\text{($M\mslabelra{\gl\pi} N$)}\\
           &  \labelra{\pi}   &\cdots (\lambda x.\Theta x \Omega)\cdots (\Omega) \cdots (\lambda x.\Theta x \Omega) \cdots&\text{($N=_{\gl\pi} \gO$)}\\
           & \mslabelra{\gl\pi}    & \cdots\cdots\cdots&
\end{array}
\]
 In order to trace the residuals it is useful to enrich the syntax of $\lambda$-terms with labels as follows:
\[
\begin{array}{lllr}
\quad M,N  & ::= & x \mid \lambda x.M \mid MN &  \\
           &     & \mid (\gl x.\Psi x\gO)^n \mid (\Psi M\gO)^n \mid (\gO)^n & (n\geq 1 \text{ and } \Psi\in \cG_\beta(\Theta)) \\
\end{array}
\]
We denote by $\Lambda^\nat$ the set of labelled terms and we write $\underline{M}$ for the $\lambda$-term, called \emph{erasure} of $M$, obtained by erasing the labels from $M$.

Since $(\gO)^n$ and $(\gl x.\Psi x\gO)^n$ are closed terms, then it is sufficient to extend substitution to labelled terms by setting $(\Psi M\gO)^n[N/x] = (\Psi M[N/x] \gO)^n$,
 where $\Psi\in \cG_\beta(\Theta)$. Then we define a  reduction on labeled terms as the smallest contextual reduction $\labelra{\mathrm{lab}}$  (reduction under labels is allowed) satisfying the following clauses, for all labelled terms $M,N$:
\[
\begin{array}{ll}
(\lambda x.M)N \labelra{\mathrm{lab}} M[N/x] & \\
\lambda x.Mx \labelra{\mathrm{lab}} M & \text{ if } x\not\in \mathrm{FV}(M) \\
(\lambda x.\Psi x\gO)^n M \labelra{\mathrm{lab}} (\Psi M\gO)^n & \text{ if } \Psi\in \cG_\beta(\Theta) \\
\Psi MN \labelra{\mathrm{lab}} \gO & \text{ if $\Psi\in \cG_\beta(\Theta)$ and $\underline{M} =_{\lambda\pi} \underline{N}$.}
\end{array}
\]
Note that (i) $\labelra{\beta\eta\pi} \subseteq \labelra{\mathrm{lab}}$; (ii) if $M\labelra{\mathrm{lab}} N$ then $(\Psi M\gO)^n \labelra{\mathrm{lab}} (\Psi N \gO)^n$.  If $\gs$ is a reduction path of labelled terms, then we denote by
 $\underline \gs$ the corresponding reduction path, where all labels are erased.

We will make use of an additional operation on labelled terms. Given terms $M,N \in \Lambda^\nat$  such that
 $\underline{M} \equiv \underline{N}$, we define their \emph{superposition} as the labelled term obtained from the syntax tree of $\underline M$ 
 by adding a possible label $k$ to each subtree $T$ of $\underline M$ according to the following schema:
\begin{itemize}
 \item Put $k = m+n$ if $T$ has label $m$ in $M$ and $n$ in $N$;
 \item Put $k=m$ if $T$ has label $m$ in $M$ and no label in $N$;
 \item Put $k = n$ if $T$ has label  $n$ in $N$ and no label in $M$;
 \item Put no label otherwise.
\end{itemize}

\begin{lem}\label{lem:proof-subst}
The term $\lambda x.\Theta x \Omega$ is $\lambda\pi$-proof-substitutable by $\ssi$.
\end{lem}

\proof
In this proof $\Psi$ ranges over $\cG_\beta(\Theta)$. Let $F_1,F_2$ be closed $\lambda$-terms and suppose $F_1(\lambda x.\Theta x \Omega) =_{\lambda\pi} F_2(\lambda x.\Theta x \Omega)$. Since the reduction
 $\labelra{\beta\eta\pi}$ is confluent, then the two sides of the equality are the beginning of two reduction paths $\gs_1$ and $\gs_2$ that end in a common term $R$. 
 Consider now the labelled terms $F_i(\lambda x.\Theta x \Omega)^{i}$ for $i=1,2$. Then there exists a labelled reduction path $\gs'_i$ starting with $F_i(\lambda x.\Theta x \Omega)^{i}$ such that $\underline{\gs'_i} \equiv \gs_i$. Notice that the  label $1$ is the unique label occurring in the reduction path $\gs'_1$, while  the label 2 is the unique label occurring in the reduction path $\gs'_2$.
 We denote by $R_i$ the last labelled term in the reduction path $\gs'_i$. Then we have that $R\equiv \underline R_1\equiv \underline R_2$.

Let $P$ be the term obtained by superposition of $R_1$ and $R_2$. Then the labels of $P$ range over the set $\cL=\{ 1,2,3\}$.
We now describe how to extract a witness of $\lambda\pi$-proof-substitutability by suitably modifying $P$. All residuals with label $3$ in $P$ are common to the reduction paths $\gs'_1$ and $\gs'_2$. Then, if we mimic the reduction path $\gs_i$  starting from $F_i\ssi$ ($i=1,2$),  we will find in place of the residuals with label $3$ 
the term $\ssi$ for $(\gl x.\Psi x \gO)^{3}$; $M$ for $(\Psi M \gO)^{3}$ and a term $N$ ($\gl\pi$-convertible with $\gO$) for $(\gO)^{3}$:
 \[
\begin{array}{lll}
F_i(\lambda x.\Theta x \Omega) &\mslabelra{\gl\pi} & \text{($i = 1,2$)} \\
\cdots (\lambda x.\Theta x \Omega)\cdots  &\mslabelra{\gb} &   \\
    \cdots (\lambda x.\Psi x \Omega)\cdots       &  \mslabelra{\gl\pi}   &\text{($\Theta\mslabelra\gb \Psi$)} \\
 \cdots (\lambda x.\Psi x \Omega)M\cdots      &  \labelra{\gb}   & \\
  \cdots \Psi M \Omega\cdots          &  \labelra{\gl\pi}   &\text{($M\mslabelra{\gl\pi} N$)} \\
  \cdots \Psi N \Omega\cdots         &  \labelra{\gl\pi}   &\text{($N=_{\gl\pi} \gO$)}\\
\cdots \Omega \cdots            &  &\\
\end{array}\qquad\qquad
\begin{array}{lll}
F_i\ssi &\mslabelra{\gl\pi} & \text{($i = 1,2$)} \\
\cdots \ssi\cdots  &\equiv &   \\
    \cdots \ssi\cdots       &  \mslabelra{\gl\pi}   & \\
 \cdots \ssi M\cdots      &  \labelra{\gb}   & \\
  \cdots M\cdots          &  \labelra{\gl\pi}   &\text{($M\mslabelra{\gl\pi} N$)} \\
  \cdots N\cdots         &  \equiv   &\text{($N=_{\gl\pi} \gO$)}\\
\cdots N \cdots            &  &\\
\end{array}
\]
Then we substitute all occurrences of the label $3$ in the term $P$ as follows: 
$$Q \equiv P[\ssi/(\gl x.\Psi x \gO)^{3}; M/(\Psi M \gO)^{3}; \gO/(\gO)^{3}].$$ 
The last substitution $\gO$ for $(\gO)^3$ is possible because the term $N$ in the above reduction path (right column) is $\gl\pi$-convertible with $\gO$. 
 We see that, by mimicking the steps in the paths $\gs_1,\gs_2$, we have that
$$(\ast)\qquad F_i\ssi =_{\gl\gp} L_i \text{, where $L_i$ is the erasure of } Q[\ssi/(\gl x.\Psi x \gO)^{i}; M/(\Psi M \gO)^{i}; \gO/(\gO)^{i}] \quad (i=1,2) $$
Let $x_1,x_2$ be fresh variables and let $H$ be the term without labels obtained from $Q$ by replacing bottom-up the subterms
$$
\text{ for $i=1,2$ }
\begin{cases}
(\gl x.\Psi x \gO)^i & \text{ with $x_i$;}   \\
(\Psi M \gO)^i       & \text{ with $x_iM$;}  \\
(\gO)^i              & \text{ with $x_i\gO$.}
\end{cases}
$$
Then the following equivalences hold:
\begin{enumerate}[label=(\alph*)]
\item $L_1 =_{\gl\gp} H[\ssi/x_1; (\gl x.\Psi x \gO)/x_2]$;
\item $L_2 =_{\gl\gp} H[(\gl x.\Psi x \gO)/x_1;\ssi/x_2]$.
\end{enumerate}
Therefore by setting $G\equiv \gl x_2x_1.H$, we obtain that
\begin{itemize}
\item $G(\gl x.\Theta x\gO)\ssi \mslabelra{\beta} H[\ssi/x_1; (\gl x.\Psi x \gO)/x_2] =_{\gl\gp} L_1 = _{\gl\gp} F_1\ssi$, \qquad by ($\ast$) and (a)
\item $G\ssi(\gl x.\Theta x\gO) \mslabelra{\beta} H[(\gl x.\Psi x \gO)/x_1;\ssi/x_2] =_{\gl\gp} L_2 =_{\gl\gp} F_2\ssi$, \qquad by ($\ast$) and (b)
\end{itemize}
This shows that $G$ is the witness term we were looking for.
\qed

Now we are ready to give the main theorem of the section.

\begin{thm}\label{thm:reduce-incons-lambdaphipi}
 The  $\lambda$-theory $\lambda\pi\phi$ is consistent.
\end{thm}

\proof 
Lemma \ref{lem:proof-subst} and Lemma \ref{lem:bottom-preorder} show that 
 the hypotheses of Theorem \ref{thm:reduce-incons} are satisfied by the equation that 
 axiomatizes $\lambda\pi\phi$ over $\lambda\pi$, and therefore $\lambda\pi\phi$ must be consistent.
\qed

%%%%%%%%%%%%%%%%%%%%%%%%%%%%%%%%%%%%%%%%%%%%%%%%%%%%%%%%%%%%%%%%%%%%%%%%%%%%%%%%%%%%%%%%%%%%%%%%%%%%%%%%%%%%%%%%%%%%%%%%%%%%%%%%%%%%%%
\section{On the equational inconsistency of the pointed ordered models}\label{sec:Hon-Plo-question}
%%%%%%%%%%%%%%%%%%%%%%%%%%%%%%%%%%%%%%%%%%%%%%%%%%%%%%%%%%%%%%%%%%%%%%%%%%%%%%%%%%%%%%%%%%%%%%%%%%%%%%%%%%%%%%%%%%%%%%%%%%%%%%%%%%%%%%

 In this section we find a counterexample to the equational consistency of the class of pointed ordered models:
we prove that there is no pointed ordered model satisfying the equations ($\pi$) and ($\phi$).

\begin{lem}\label{lem:connected-components}
Let $\cM$ be an ordered model such that $\cM\models \Theta xx=\gO \land \Theta x\gO = x$ (i.e., $Th(\cM) \supseteq \gl\pi\phi$). Then for all closed $\gl$-terms $P$ and $Q$ we have:
\begin{enumerate}[label=(\roman*)]
\item If  $\cM \not\models\Theta PQ= \gO$, then the interpretations of $P$ and $Q$ are in distinct connected components of $\cM$.
\item The connected component of the interpretation of  $\gO$ is a singleton set.
\end{enumerate}
\end{lem}

\proof (i) Following \cite[Section 4]{Salibra03} we define the subtraction sequence of the pair $(P,Q)$: 
$$s_1 \equiv \Theta PQ; \qquad s_{n+1} \equiv \Theta s_n \gO.$$
 By hypothesis  $\cM\models s_1 \neq\Omega$ and by subtractivity   $\cM\models s_n  = s_1$ for all $n$.
Then the conclusion follows from \cite[Corollary 4.6]{Salibra03}.

 (ii) Since $\cM$ is a subtractive ordered model then the conclusion follows from Lemma \ref{lem-ord}.
\qed

The situation described by Lemma \ref{lem:connected-components} can be regarded to as a relativized version of absolute unorderability to one fixed element.
 In particular the interpretation of $\Omega$ is isolated in every model. This property will be studied in Section \ref{sec:subtractivity}
 in the framework of Universal Algebra.

The class of pointed ordered models is not consistent with respect to the set of quantifier-free sentences: the quantifier-free sentence
 $\gl x.\gO xx = \gl x.\gO \land \gO\neq\gO\gO(\gO\ssT\ssi)$ is consistent with the $\gl$-calculus but no pointed ordered model satisfies
 it. This result was shown by Salibra for the $\gl$-calculus (see the remark after \cite[Corollary 3.6]{Salibra03}) and by
 Honsell\textendash Plotkin for the extensional $\gl$-calculus (see \cite[Theorem 7]{HonsellP09}): Honsell and Plotkin also ask
 whether there exist a finite set of equations consistent with the $\gl$-calculus that are not satisfied by any pointed ordered model.
 The following theorem answer their question: the subtractive equations are indeed a counterexample to the equational consistency for the class of pointed ordered models.

\begin{thm}
 No pointed ordered model $\cM$ simultaneously  satisfies  the equations $\Theta xx=\gO\ \land\ \Theta x\gO = x$.
\end{thm}

\proof
Suppose, by contradiction, that $\cM\models \Theta xx=\gO \land  \Theta x\gO = x$.
 Since $|\gO|^{\cM}$ is comparable with $\bot$, then by Lemma \ref{lem:connected-components}(ii) $\gO$ is interpreted as the bottom
 element $\bot$ of $\cM$.  Since the bottom element is comparable with all other elements of $\cM$, from Lemma \ref{lem:connected-components}(ii) it follows that $\cM$ is trivial.
\qed

 We can also get a stronger result.

\begin{thm} 
 No connected ordered model  simultaneously  satisfies the equations $\Theta xx=\gO\ \land\ \Theta x\gO = x$.
\end{thm}

\proof
By Lemma \ref{lem:connected-components}(ii).
\qed

%%%%%%%%%%%%%%%%%%%%%%%%%%%%%%%%%%%%%%%%%%%%%%%%%%%%
\section{On the order-incompleteness of $\gl$-calculus}\label{subs:strenghtening}
%%%%%%%%%%%%%%%%%%%%%%%%%%%%%%%%%%%%%%%%%%%%%%%%%%%

 The \emph{order-incompleteness} problem of $\gl$-calculus, raised by Selinger in \cite{Selinger03}, can be  characterised in terms of connected components  of a partial ordering: 
a  $\gl$-theory $\cT$ is order-incomplete if, and only if, every ordered model $\cM$ such that $Th(\cM) = \cT$ is partitioned in an infinite number of connected components, each one containing exactly one element. In other words, the partial order is the equality. 

So far we have shown that the subtractive equations force the connected component of $\gO$ in a model to be a singleton set, so that the model cannot be connected as a partial order.  However, the
 order-incompleteness is something more than disconnected. Toward order-incompleteness, we propose a strengthening $\cT$ of the
 $\gl$-theory $\lambda\pi\phi$ having the following property: every ordered model $\cM$ such that $Th(\cM) \supseteq \cT$ has an infinite
 number of connected components among which that of $\gO$ is a singleton set. Moreover each connected component contains the
 denotation of at most one $\beta\eta$-normal form.

We define a family of unsolvable terms $\Theta_n$ (for $n\geq 0$) obtained as follows:
\begin{itemize}
\item Define inductively $A_0 \equiv x$ and $A_{n+1} \equiv \lambda y.yA_n$, where $y\not\equiv x$. Note that $\mathrm{FV}(A_n) = \{x\}$, for each $n\in\nat$.
\item Now set $B_n \equiv \lambda x.xA_n$, $C_n \equiv \lambda z.zB_n$ (where $z\not\equiv x$) and $\Theta_n \equiv B_nC_n$.
\end{itemize}
 Note  that $\Theta_0 \mslabelra{\beta} \Omega$ and
 $\Theta_1 \equiv \Theta$ (recall the definition of $\Theta$ from Section \ref{sus:lam-calc}).

\begin{lem}
The terms $\Theta_n$ are closed zero-terms such that
  $\lambda\pi\vdash\Theta_n = \Theta_m$ iff $m=n$.
\end{lem}

\proof
It follows from the confluence of the reduction
 $\labelra{\beta\eta\pi}$ and from the form of the reduction graphs of the terms in question. Each graph $\cG_{\beta\eta\pi}(\Theta_n)$
 is a cycle whose edges are only $\labelra{\beta}$ reductions, and $\cG_{\beta\eta\pi}(\Theta_n)$ is disjoint from $\cG_{\beta\eta\pi}(\Theta_m)$
 whenever $n\neq m$. 
\qed

We are now going to introduce the above-mentioned strenghtening of $\lambda\pi\phi$. In what follows we let $\cT$ be the theory axiomatized over
 $\lambda\pi\phi$ by the following equations:
\[
\begin{array}{ll}
\Theta_2\gO = \ssT; & \\
\Theta_2(\Theta MN) = \ssF, &\quad \text{$M$ and $N$ distinct closed $\beta\eta$-normal forms.}
\end{array}
\]
(recall the definiton of $\ssT$ and $\ssF$ from Section \ref{sus:lam-calc}.)

Next we show that $\cT$ is consistent. In order to do that it suffices, by compactness reasons, to prove that any finite subset of the above equations
 is eliminable from a proof of $\cT \vdash \ssT = \ssF$ via the Jacopini\textendash Kuper technique. 
 The proof of this fact closely resembles the consistency proof given for $\gl\pi\phi$ (see Section \ref{subs:Jac-for-lpp}), so
 we will just sketch it, only considering the extension of $\gl\pi$ by three equations $\Theta_2(\Theta MN) = \ssF$, $\Theta_2\gO = \ssT$ and 
 $\lambda x.\Theta x\Omega = \ssi$, where $(M,N)$ is an arbitrary but fixed pair of closed distinct $\beta\eta$-normal forms.

Define the two sequences $\vec P = \Theta_2(\Theta MN),\Theta_2\gO,\lambda x.\Theta x\Omega$ and $\vec Q = \ssF,\ssT,\ssi$.

\begin{lem}  $\vec P$ is $\lambda\pi$-operationally less defined than $\vec Q$. 
\end{lem}

\proof
As the proof of Lemma \ref{lem:bottom-preorder}.
\qed

\begin{lem}\label{lem:proof-rep-2}
$\vec P$ is $\lambda\pi$-proof-substitutable by $\vec Q$.
\end{lem}

\proof
In this proof $\Psi$ and $\Psi_2$ range, respectively, over $\cG_\beta(\Theta)$ and $\cG_\beta(\Theta_2)$.
 Let $F_1,F_2$ be closed $\lambda$-terms and suppose $F_1\vec P =_{\lambda\pi} F_2\vec P$. Since the reduction
 $\labelra{\beta\eta\pi}$ is confluent, then the two sides of the equality are the beginning of two reduction paths $\gs_1$ and $\gs_2$ that end in a common term $R$. 

Consider now the labelled term
\begin{itemize}
\item $A_1 \equiv F_1(\Theta_2(\Theta MN))^{1}(\Theta_2\gO)^{4}(\lambda x.\Theta x \Omega)^{10}$
\item $A_2 \equiv F_2(\Theta_2(\Theta MN))^{2}(\Theta_2\gO)^{5}(\lambda x.\Theta x \Omega)^{11}$
\end{itemize}
Then there exist labelled reduction paths $\gs'_i$ starting with $A_i$ ($i=1,2$) such that $\underline{\gs'_i} \equiv \gs_i$.
 We denote by $R_i$ the last labelled term in the reduction path $\gs'_i$. Then we have $R\equiv \underline R_i$ ($i = 1,2$).
 Let $S$ be the term obtained by superposition of $R_1$ and $R_2$. Then the labels of $S$ range over the set $\cL=\{1,2,3,4,5,9,10,11\}$. Note that if $S$ has a labelled subterm of the shape $(\Theta_2\gO)^{l}$, then $l \in \{4,5,9\}$ 
 because the contrary would require $\Theta MN \labelra{\beta\eta\pi} \gO$ (by Theorem \ref{thm:betapil-confluent}(ii)), which is impossible because it would imply
 $\lambda\pi \vdash M = N$, contradicting the consistency of $\lambda\pi$ (as a consequence of B\"{o}hm's Theorem \cite[Thm.~10.4.2]{Bare}).

We now describe how to extract a witness of $\lambda\pi$-proof-substitutability by suitably modifying $S$. All residuals with label $3$ ,$9$, or $21$ in $S$
 are common to the reduction paths $\gs'_1$ and $\gs'_2$. Then, if we mimic the reduction path $\gs_i$ starting from $F_i\ssi$ ($i=1,2$), we will find in place
 of the residuals with label $21$ the term $\ssi$ for $(\gl x.\Psi x \gO)^{21}$; $M$ for $(\Psi M \gO)^{21}$ and a term $N$ ($\gl\pi$-convertible with $\gO$) for $(\gO)^{21}$.
 Similarly those residuals with labels $3$ and $9$ are replaced by $\ssF$ and $\ssT$, respectively. Then we let 
$$S' \equiv S[\ssi/(\gl x.\Psi x \gO)^{21}; M/(\Psi M \gO)^{21}; \gO/(\gO)^{21};\ssF/(\Theta_2(\Theta MN))^{3};\ssT/(\Theta_2\gO)^{9}]$$
and we define a term $H$ out of $S'$ by replacing bottom-up some subterms (labeled by $i\in \cL$), using fresh variables $x_{1},x_{2},x_3,x_4,x_5,x_{10},x_{11}$ as follows
$$
\text{ for $i=10,11$ }
\begin{cases}
(\gl x.\Psi x \gO)^i & \text{with $x_i$;}   \\
(\Psi M \gO)^i       & \text{with $x_iM$;} \\
(\gO)^i              & \text{with $x_i\gO$.} 
\end{cases}
\quad
\text{for $i=4,5$ and $j=1,2$}
\begin{cases}
(\Psi_2\gO)^i       & \text{with $x_i$;}   \\
(\Psi_2(\Psi MN))^j & \text{with $x_j$.}
\end{cases}
$$
Finally, as in the proof of Lemma \ref{lem:proof-subst}, it is possible to find a term $G$ such that:
\[
\begin{array}{rclcl}
G\vec P \vec Q & \mslabelra{\beta} & H[(\Psi_2(\Psi MN)) / x_1; \ssF / x_{2} ; (\Psi_2\gO)/ x_{4}; \ssT /x_5; \ssi / x_{10}; (\gl x.\Psi x \gO)/ x_{11}]
               & =_{\gl\gp}                 & F_1\vec Q \\
               & & & & \\
G\vec Q \vec P & \mslabelra{\beta} & H[\ssF / x_1;(\Psi_2(\Psi MN)) / x_{2} ;\ssT / x_{4}; (\Psi_2\gO)/x_5; (\gl x.\Psi x \gO)/ x_{10}; \ssi/ x_{11}]
               & =_{\gl\gp}                 & F_2\vec Q \\
\end{array}
\]
\qed

The following proposition, which relies on Lemma \ref{lem:proof-rep-2}, it is analogous to Theorem \ref{thm:reduce-incons-lambdaphipi}.

\begin{thm}\label{thm:super-cool}
The $\gl$-theory $\cT$ is consistent.
\end{thm}

In \cite{Salibra03} it is shown that the semantics of $\gl$-calculus given in terms of ordered models with a finite number of connected components is theory incomplete. In Theorem \ref{thm:order-inc} below we improve this result.

\begin{lem}\label{lemma:infinite-components} 
Let $\cM$ be an  ordered model such that $Th(\cM) \supseteq \cT$. Then $\cM$ has an infinite number of connected components and it has the following properties:
\begin{enumerate}[label=(\roman*)]
\item The interpretation in $\cM$ of two distinct $\beta\eta$-normal forms belongs to different connected components;
\item The connected component of $\gO$ is a singleton set.
\end{enumerate}
\end{lem}

\proof
Let $M,N$ be two distinct $\beta\eta$-normal forms and suppose, by way of contradiction, that $M$ and $N$ lie in the same connected component
 of $\cM$. Then $\cM \models \Theta MN = \gO$ by Lemma \ref{lem:connected-components}(i).
 From $\cM \models \ssF = \Theta_2(\Theta MN)$ and $\cM \models \Theta_2\gO = \ssT$ we derive that $\cM \models \ssF = \ssT$,
 which contradicts the non-triviality of $\cM$. Hence each denotation of a $\beta\eta$-normal form belongs to exactly one connected component.
The second part of the statement follows directly from Lemma \ref{lem:connected-components}(ii).
\qed

\begin{thm}\label{thm:order-inc} The following classes of ordered models are theory-incomplete:
\begin{enumerate}[label=(\roman*)]
\item Models with a finite number of connected components.
\item Models with an infinite number of connected components such that the connected component of $\gO$ has cardinality $>1$.
\end{enumerate}
\end{thm}

\proof By Lemma \ref{lemma:infinite-components}.
\qed

%%%%%%%%%%%%%%%%%%%%%%%%%%%%%%%%%%%%%%%%%%%%%%%%%
\section{Subtractivity and orderings}\label{sec:subtractivity}
%%%%%%%%%%%%%%%%%%%%%%%%%%%%%%%%%%%%%%%%%%%%%%%%

The inspiration for the subtractive equations  comes from a general algebraic framework, developed by Ursini \cite{Ursini}, called \emph{subtractivity}.  Salibra in \cite{Salibra03} investigated the weaker notion of \emph{semi-subtractivity}, linking it to properties of ordered models of $\lambda$-calculus.
 Here we follow that path illustrating the stronger properties of subtractivity.

We start the section reviewing the connection
 established by Selinger in \cite{Selinger03} between the absolute unorderability and the validity of certain Mal'cev-type conditions.

\subsection{Unorderability and absolute unorderability}

Let $\gt$ be an algebraic similarity type and $\bA$ be an algebra of type $\gt$. 
We say that $\bA$ is \emph{unorderable} if it admits only equality as a compatible partial order.

The following result is due to Hagemann \cite{Hag73,HM73} (see also Coleman \cite[Theorem 1.6]{Coleman97}).

\begin{thm}\label{thm-unord} Let $\cV$ be a variety. Then the following conditions are equivalent:
\begin{enumerate}
\item Every algebra in $\cV$ is unorderable;
\item Every  compatible preorder on an algebra in $\cV$ is symmetric (and thus a congruence).
\item There exist a natural number $n\geq 2$ and ternary terms $p_1,\ldots,p_{n-1}$
 in the type of $\cV$ such that the following Mal'cev identities hold in $\cV$:
  \vspace*{-1.6mm}
 $$
 \begin{array}{rcl}
          x & = & p_1(x,y,y); \\
 p_i(x,x,y) & = & p_{i+1}(x,y,y) \quad (i=1,\dots,n-2); \\
 p_{n-1}(x,x,y) & = & y.
 \end{array}
 $$
\end{enumerate}
\end{thm}

\proof The equivalence of (2) and (3) is \cite[Theorem 1.6]{Coleman97}. We now prove that (1) implies (2).
Let $\bA\in\cV$ and $\leq$ be a compatible preorder on $\bA$.  
 Let $\approx$ be the congruence on $\bA$ generated by $\leq$, that is, 
$$\text{$a \approx b$ iff $a\leq b$ and $b\leq a\quad$ (for all $a,b\in A$).}$$
By hypothesis the partial ordering on the quotient algebra $\bA/\approx$, defined by
$[a]_\approx \sqsubseteq [b]_\approx$ iff $a\leq b$, is trivial.
Then $\leq$ is symmetric.
\qed

\begin{defi}
Let $\Cc$ be a class of algebras of type $\gt$ and $\bA\in \Cc$. 
We say that $\bA$ is \emph{absolutely unorderable in $\Cc$} if, for every algebra $\bB\in \Cc$ and 
 every embedding $f: \bA \to \bB$ (i.e., injective homomorphism), the algebra $\bB$ is unorderable.
\end{defi}

Let $\cV$ be a variety of type $\gt$ and $\bA\in \cV$. We denote by $\gt(A)$  the algebraic similarity
type $\gt$ enriched by a constant $\bar a$ 
for each element $a\in A$. The algebra $\bA$ becomes a $\gt(A)$-algebra by interpreting each constant $\bar a$
with the element $a\in A$.
The  {\it equational diagram\/} of $\bA$ is defined as
the set of all \emph{ground identities} $t = u$ of type $\gt(A)$ such that $\bA\models t=u$.

 We denote by $\cV_\bA$ the variety of type $\gt(A)$ axiomatized by the equational theory $Eq(\cV)$ of $\cV$
 and the equational diagram of $\bA$.

There is a bijective correspondence between algebras of $\cV_\bA$ and pairs $(\bB,f)$, where
 $\bB\in\cV$ and $f:\bA\to\bB$ is
a $\gt$-homomorphism. Indeed,
the algebra $\bB$ can be turned into a $\gt(A)$-algebra in $\cV_\bA$, by interpreting each constant $\bar a$ with the element $f(a)$ of $B$.

If $X$ is a set of indeterminates, then  the {\it free extension $\bA[X]$  of $\bA$ by $X$} in the
variety  $\cV$ is  the free $\gt(A)$-algebra over $X$
in the variety $\cV_\bA$.
The algebra $\bA[X]$  can be also defined up to isomorphism by the following universal mapping properties: (1) $A\cup X\subseteq A[X]$;  (2) $\bA[X]\in \cV$; (3) for every  $\bB\in \cV$,  $\gt$-homomorphism $h:\bA\to\bB$ and every function $f:X\to B$, there exists a unique $\gt$-homomorphism $g:\bA[X]\to \bB$ extending $h$ and $f$.
When $X= \{x_1,\dots, x_n\}$ is finite, we write $\bA[x_1,\dots, x_n]$ for $\bA[X]$.

We are now ready to give the main result of this subsection that characterizes those algebras which are absolutely unorderable in a variety. Notice that the equivalence of items (1) and (3) below was shown by  Selinger in \cite[Theorem 3.4]{Selinger03}, while the equivalence of (1) and (2) was suggested by a referee.

\begin{thm}\label{ab-thm} Let $\cV$ be a variety of type $\gt$ and $\bA\in \cV$. Then the following conditions are equivalent:
\begin{enumerate}
\item $\bA$ is absolutely unorderable in $\cV$;
\item  Every algebra in the variety $\cV_\bA$ is unorderable;
\item The algebra $\bA[x,y]$ satisfies the Mal'cev identities of Theorem \ref{thm-unord}.
\end{enumerate}
\end{thm}

\proof (1) $\Rightarrow$ (2) Assume, by contradiction, $\bB$ to be a $\gt(A)$-algebra in $\cV_\bA$, which admits a nontrivial compatible partial order $\leq$. 
Then there are two distinct elements $a,b\in B$ such that $a\leq b$. 
 If $f:X\to B$ is an onto map from a set  $X$ of indeterminates  to $B$, then there are two indeterminates $x,y\in X$ such that $f(x)= a$ and $f(y)=b$. Let $\theta$ be the least compatible preorder on $\bA[X]$ such that 
 $x\theta y$, and
 let $\approx$ be the congruence on $\bA[X]$ generated by $\theta$, that is, 
$$\text{$t \approx u$ iff $t\theta u$ and $u\theta t\quad$ (for all $t,u\in \bA[X]$).}$$
Then the relation $\sqsubseteq$, defined by  $[t]_{\approx} \sqsubseteq [u]_{\approx}$ iff $t\theta u$, is a compatible partial ordering on $\bA[X]/\!\!\approx$.
 By the hypothesis $x\theta y$ we obtain $[x]_{\approx}\sqsubseteq [y]_{\approx}$. Since the map $\iota: \bA\to \bA[X]/\!\!\approx$, defined by $\iota(a) = [a]_{\approx}$, is an embedding and $\bA$  is absolutely unorderable in $\cV$, then 
 we  get $[y]_{\approx}= [x]_{\approx}$, so that $y\theta x$ holds.
 Since $\bB\in \cV_\bA$, then the map $h:A\to B$, defined by $h(a)= {\bar a}^\bB$, is a $\gt$-homomorphism. Consider the
unique $\gt$-homomorphism $g:\bA[X]\to \bB$ extending $h$ and $f$.
 Since $\theta \subseteq \{(t,u) : g(t) \leq g(u)\}$, then by $y\theta x$ we get $b=g(y) \leq g(x)=a$, that together with $a\leq b$ implies $a=b$. Contradiction.

 (2) $\Leftarrow$ (1)  If $\bB\in\cV$ and  $f: \bA\to\bB$ is an embedding, then $\bB$ can be seen as a $\gt(A)$-algebra in $\cV_\bA$, which is unorderable by hypothesis.
 
 (2) $\Leftrightarrow$ (3) follows from Theorem \ref{thm-unord}.
\qed

It is unknown whether there exist absolutely unorderable algebras in the variety of combinatory algebras. Plotkin and Simpson have shown that the Mal'cev  
identities are inconsistent with combinatory logic for $n=2$, while Plotkin and Selinger have obtained the same result for $n=3$ (see \cite{Selinger03}). It is unknown whether the Mal'cev identities are consistent with combinatory logic for $n\geq 4$.

In \cite{Salibra03} the second author has shown that there exists a quasi-variety of combinatory algebras which only admits absolutely unorderable algebras. 
We recall that a quasi-variety is a class of algebras axiomatized by quasi-identities (i.e., equational implications with a finite number of equational
premises). 

\begin{thm} \rm{(\cite{Salibra03})} Let $\cQ$ be the quasi-variety of combinatory algebras axiomatized by the identity $\gO xx=\gO$ and the quasi-identity $\gO xy=\gO \Rightarrow x=y$. Then  $\cQ$ is nontrivial and every algebra $\bA\in\cQ$ is absolutely unorderable in $\cQ$.
\end{thm}

\proof
$\cQ$ is nontrivial because the term model of the consistent $\gl$-theory axiomatised by $\gO xx=\gO$ belongs to $\cQ$. Moreover, it is easy to show that every algebra  of $\cQ$ is unorderable.
\qed

%%%%%%%%%%%%%%%%%%%%%%%%%%%%%%%%%%%%%%%%%%%%%%%%%%%%%%%%%%%%%%%%%%%%
\subsection{Absolute $0$-unorderability}
%%%%%%%%%%%%%%%%%%%%%%%%%%%%%%%%%%%%%%%%%%%%%%%%%%%%%%%%%%%%%%%%%%%%

In the case a variety $\cV$ has two constants $0$ and $1$, the Mal'cev identities of Theorem \ref{thm-unord} give:
  $$
 \begin{array}{rcl}
          0 & = & p_1(0,1,1); \\
 p_i(0,0,1) & = & p_{i+1}(0,1,1) \quad (i=1,\dots,n-2); \\
 p_{n-1}(0,0,1) & = & 1.
 \end{array}
 $$
If we define the unary term operations $f_i(x)= p_i(0,x,1)$, then the above identities can be written as follows:

\begin{equation}\label{eq}
0= f_1(1);\qquad f_i(0)  =  f_{i+1}(1) \quad (i=1,\dots,n-2);\qquad  f_{n-1}(0)  =  1.
\end{equation}

This suggests the following theorem, whose proof is similar to the proof of Theorem \ref{thm-0-unord} below and it is omitted.

\begin{thm}
 Let $\cV$ be a variety with two constants $0$ and $1$. Then the
  constants $0$ and $1$ are incomparable in all ordered algebras in $\cV$ if, and only if,
  there exist a natural number $n\geq 2$ and unary terms $f_1,\ldots,f_{n-1}$
 in the type of $\cV$ such that the identities (\ref{eq}) hold in $\cV$.
\end{thm}

In the case a variety $\cV$ has a constant $0$, then we can relativise the Mal'cev identities of Theorem \ref{thm-unord} as follows:
  $$
 \begin{array}{rcl}
          0 & = & p_1(0,y,y); \\
 p_i(0,0,y) & = & p_{i+1}(0,y,y) \quad (i=1,\dots,n-2); \\
 p_{n-1}(0,0,y) & = & y.
 \end{array}
 $$
 If we define the binary term operations $s_i(y,x)= p_i(0,x,y)$, then the above identities can be written as follows:
 \begin{equation}\label{eq2}
\begin{array}{rcl}
           0 & = & s_1(x,x) \\
    s_i(x,0) & = & s_{i+1}(x,x) \quad (i=1,\dots,n-2); \\
s_{n-1}(x,0) & = & x.  
\end{array}
\end{equation}

This suggests that the absolute unorderability relative to the element $0$ can be expressed by the following identities defining $n$-subtractivity.

\begin{defi} 
 An algebra $\bA$ is \emph{$n$-subtractive} ($n\geq 2$) if there exist a constant $0$ and $n-1$ binary terms $s_1(x,y),\dots, s_{n-1}(x,y)$ such that $\bA$ satisfies  identities (\ref{eq2}).
A variety of algebras is $n$-subtractive if every algebra in $\cV$ is $n$-subtractive with respect to the same constant and term operations.
\end{defi}

Then Ursini's subtractivity of Definition \ref{sub} means $2$-subtractivity: this is the strongest notion since it is easy to verify that
 an $n$-subtractive algebra is also $m$-subtractive for every $m > n$.

Every model of the two equations $(\pi)$ and $(\phi)$ is subtractive, when we define the binary subtractivity operator $s_1(x,y)$  as the $\gl$-term $\Theta xy$. As a consequence of the consistency of the $\gl$-theory $\gl\pi\phi$, it follows that
 there exists a non-trivial subtractive variety of combinatory algebras.

\begin{defi}
An algebra $\bA$ with $0$ is 
\begin{enumerate}[label=(\roman*)]
\item \emph{$0$-unorderable} if, for every compatible partial order $\leq$ on $A$ and  every $a\in A\setminus\{0\}$, neither $0\leq a$ nor $a\leq 0$.
\item \emph{$0$-symmetric} if, for every compatible preorder $\leq$ on $A$ and  every $a\in A\setminus\{0\}$, we have $0\leq a \Leftrightarrow a\leq 0$.
\end{enumerate}
\end{defi}

 \begin{prop}\label{subunord} Every $n$-subtractive algebra  is $0$-unorderable and $0$-symmetric.
\end{prop}

\proof
Let $\bA$ be $n$-subtractive and $\leq$ be a compatible  preorder on $A$.  If $a\in A$ and $a\leq 0$, then
$$0= s_{1}(a,a) \leq s_{1}(a,0)= s_{2}(a,a) \leq \dots\leq s_{n-2}(a,0)=s_{n-1}(a,a) \leq s_{n-1}(a,0) = a.$$
If $0\leq a$ a similar reasoning works.
\qed

The following theorem relativizes Theorem \ref{thm-unord}  to $0$-unorderability.

\begin{thm}\label{thm-0-unord} Let $\cV$ be a variety with a constant $0$. Then the following conditions are equivalent:
\begin{enumerate}
\item Every algebra in $\cV$ is $0$-unorderable;
\item Every  compatible preorder on an algebra in $\cV$ is $0$-symmetric.
\item $\cV$ is $n$-subtractive for some $n$.
\end{enumerate}
\end{thm}

\proof  (2) $\Rightarrow$ (3):  
Define a compatible relation $\prec$ on the free algebra $\bT_\cV[x]$ as follows:
 $t \prec u$ iff there exists a binary term $p(x,y)$ such that $\cV\models t = p(x,x)$ and $\cV \models p(x,0) = u$.
The condition $x \prec 0$ is witnessed by the polynomial $p(x,y) \equiv y$. The reflexive and transitive closure $\prec^*$ of $\prec$ is a compatible preorder on $\bT_\cV[x]$. Then by hypothesis we derive $0\prec^* x$. This implies the existence of binary terms $s_1,\ldots,s_{n-1}$ which witness $n$-subtractivity.

(3) $\Rightarrow$ (2): By Proposition \ref{subunord}.

(1) $\Leftrightarrow$ (2): as in Theorem \ref{thm-unord}.
\qed

\begin{defi} Let $\Cc$ be a class of algebras with a constant $0$.
An algebra $\bA\in \Cc$ is said to be \emph{absolutely $0$-unorderable in $\Cc$} if, for any algebra $\bB\in \Cc$ and embedding $f: \bA \to\bB$,  $\bB$ is $0$-unorderable.
\end{defi}

\begin{thm}\label{ab2-thm} Let $\cV$ be a variety of type $\gt$ and $\bA\in \cV$. Then the following conditions are equivalent:
\begin{enumerate}
\item $\bA$ is absolutely $0$-unorderable in $\cV$;
\item  Every algebra in the variety $\cV_\bA$ is $0$-unorderable;
\item The algebra $\bA[x]$ is $n$-subtractive for some $n$.
\end{enumerate}
\end{thm}

\proof (2) $\Leftrightarrow$ (3) follows from Theorem \ref{thm-0-unord}.

(1) $\Rightarrow$ (2): As the corresponding proof of Theorem \ref{ab-thm}, where the role of the indeterminate $y$ is taken by the constant $0$.

(2) $\Rightarrow$ (1):  If $\bB\in\cV$ and  $f: \bA\to\bB$ is an embedding, then $\bB$ can be seen as a $\gt(A)$-algebra in $\cV_\bA$, which is $0$-unorderable by hypothesis.
 \qed

\begin{cor}\label{mostro}
Every $\gl$-model satisfying the subtractive identities $\pi$ and $\phi$
is absolutely $\gO$-unorderable in the variety of combinatory algebras.
\end{cor}

 \proof Let $\bA$ be a $\gl$-model satisfying the subtractive identities $\pi$ and $\phi$.  
By \cite[Proposition 20]{Sel-alg} and \cite[Theorem 4.2]{PigozziS98}, for all $\gl$-terms $M$ and $N$, $\bA \models M=N$ if, and only if, $\bA[x]\models M = N$. Then $\bA[x]$ is also subtractive. Then the conclusion follows from Theorem \ref{ab2-thm}.
\qed

We would like to conclude this section by remarking that
Ursini \cite{Ursini} has shown that subtractive algebras have a good theory of ideals. 
We recall that ideals in general algebras generalize normal subgroups, ideals in rings, filters in Boolean or Heyting algebras, ideals in Banach algebra, in $l$-groups, etc.  One  feature of subtractive varieties is that their ideals are exactly the congruence classes of $0$, but one does not have the usual one-one correspondence ideals-congruences: mapping a congruence $\theta$ to its equivalence class $0/\theta$ only establishes a lattice homomorphism between the congruence lattice and the ideal lattice. This points to another feature: the join of two congruences is a tricky thing to deal with. The join of two ideals in a subtractive algebra behaves nicely: for $I,J$ ideals, we have that $b\in I\lor J$ iff for some $a\in I$, $s(b, a)\in J$. Thanks to the consistency of the subtractive equations with $\gl$-calculus, the theory of ideals for subtractive varieties can be applied to all $\gl$-theories extending $\gl\pi\phi$. 

%%%%%%%%%%%%%%%%%%%%%%%%%%%%%%%%%%%%%%%%%%%%%%%%%%%%%%%%%%%%%%%%%%%%
\section{Subtractivity and topology}\label{sec:subtractivity-topology}
%%%%%%%%%%%%%%%%%%%%%%%%%%%%%%%%%%%%%%%%%%%%%%%%%%%%%%%%%%%%%%%%%%%%
In this section we provide a topological incompleteness theorem for the $\gl$-calculus as a consequence of a study of conditions of separability for $n$-subtractive algebras.

The classification of the models of lambda calculus into 
orderable/unorderable models was refined as follows in \cite{Salibra03}.
For every algebra $\bA$, let $T_i^\bA$ ($i = 0, 1, 2, 2_{1/2}$) 
be the set of all topologies $\gt$ on $A$
which make $(\bA,\gt)$ a $T_i$-topological algebra.
It is obvious that in general we have
$$T_0^\bA \supseteq T_1^\bA \supseteq T_2^\bA \supseteq T_{2_{1/2}}^\bA.$$
A topology $\gt$
with a non-trivial specialization order (we have $a <_\gt b$ for some $a,b$)
would be $T_0$ yet not $T_1$, so that
$$\mbox{$\bA$ is unorderable iff $T_0^\bA = T_1^\bA$.}$$

\begin{defi}
  We say that $\bA$ is  of \emph{topological type $i$} ($i = 1, 2, 2_{1/2}$) if $T_0^{\bA} = T_i^{\bA}$.
A variety $\cV$ of algebras is of topological type $i$ if every algebra $\bA\in\cV$ is of topological type $i$.
\end{defi}

\begin{exa} We recall from \cite{Salibra03} that a lambda theory $\cT$ is {\it of topological type $i$} ($i = 1, 2, 2_{1/2}$)
if the term model $\cM_\cT$ of $\cT$ satisfies $T_0^{\cM_\cT} = T_i^{\cM_\cT}$.
The lambda theory $\cB$, generated by equating two lambda terms 
if they have the same B\"ohm tree, is not of type $1$
(see \cite{Bare}).
$\gl\gb$ and $\gl\gb\geta$ are of type 1 by Selinger's result \cite{Selinger03}, 
while the lambda theory $\gP$, generated by the equation $\gO xx=\gO$, was shown of type $2_{1/2}$ in \cite{Salibra03}.
\end{exa}

We now refine the topological axioms of separability as follows.

\begin{defi} Let $(X,\gt)$ be a space and $a\in X$.
We say that $X$ is {\it $T_i$-separated in  $a$} ($i =0,1,2, 2_{1/2}$) if, for all $b\in X\setminus \{a\}$,
$a$ and $b$ are $T_i$-separated.
\end{defi}

For every algebra $\bA$ and $a\in A$, let $T_i^\bA(a)$ ($i = 0, 1, 2, 2_{1/2}$) 
be the set of all topologies $\gt$ on $A$
which make $(\bA,\gt)$ a topological algebra which is $T_i$-separated in $a$.
We have
$$\mbox{$\bA$ is $a$-unorderable iff $T_0^\bA(a) = T_1^\bA(a)$.}$$

\begin{defi}
  We say that $\bA$ is  of \emph{topological type $i$ in $a$} ($i = 1, 2, 2_{1/2}$) if $T_0^{\bA}(a) = T_i^{\bA}(a)$.
A variety $\cV$ of algebras, whose type contains a constant $a$, is of topological type $i$ in $a$ if every algebra $\bA\in\cV$ is of topological type $i$ in $a$.
\end{defi}

We know from Proposition \ref{subunord} that the variety of combinatory algebras generated by the term model of $\gl\gp\phi$ is of topological type $1$ in  $\gO$.
This result will be improved in the next subsection.

%%%%%%%%%%%%%%%%%%%%%%%%%%%%%%%%%%%%%%%%%%%%%%%%%%%%%%%%%%%%%%%%%%%%%%%%%%%
\subsection{The topological incompleteness theorem for $\gl$-calculus}
%%%%%%%%%%%%%%%%%%%%%%%%%%%%%%%%%%%%%%%%%%%%%%%%%%%%%%%%%%%%%%%%%%%%%%%%%%%

The \emph{$i$-diagonal} $Diag_i(\bA)$ of an $n$-subtractive algebra $\bA$ is the set of elements $a\in A$ such that 
$s_i(a,a) =0$.
Notice that (i) $Diag_1(\bA) = A$; (ii) $0\in Diag_i(\bA)$ for all $i$; (iii) $Diag_i(\bA) \supseteq Diag_{n-1}(\bA)$ for every $i$.

\begin{lem}\label{separated} Every $n$-subtractive $T_0$-semitopological algebra $(\bA,\gt)$ is $T_1$-separated in the element $0$.
\end{lem}

\proof 
By Lemma \ref{1.3.1bis} the specialisation order $\leq_\gt$ is compatible. Then the conclusion follows from Proposition \ref{subunord}. 
\qed

\begin{cor}\label{closed} 
Let $(\bA,\gt)$ be an $n$-subtractive $T_0$-semitopological algebra. Then we have:
\begin{enumerate}[label=(\roman*)]
\item The singleton set  $\{0\}$ is closed;
\item The sets $Diag_i(\bA)$ ($1\leq i\leq n-1$) are  closed.
\end{enumerate}
\end{cor}

\proof (i) An element $a$ belongs to the closure of $\{0\}$ if, and only if, $a\leq_\gt 0$. By Proposition \ref{subunord} $\bA$ is $0$-unorderable. Then the set $\{0\}$ is closed.  

(ii) $Diag_i(\bA)$ is the inverse image of the closed set $0$ with respect to the continuous unary polynomial $s_i(x,x)$.
\qed

\begin{lem}\label{semi} Let $(\bA,\gt)$ be a semitopological algebra in an arbitrary similarity type, $t(x,y)$ be a binary term operation and $a,b\in A$ be two elements such that 
$$t(a,a)=t(b,b);\qquad t(a,b)\neq t(a,a).$$  
If $t(a,b)$ and $t(a,a)$ are $T_0$-separated, then $a$ and $b$ are $T_1$-separated.
\end{lem}

\proof  Let $c\equiv t(a,a)$ in this proof.
Assume first there exists a neighbourhood $U$ of $c$ with
$t(a,b)\notin U$.
From $t(a,a) = c \in U$ and $t(b,b) = c \in U$ and from the
separated continuity of $t$ it follows that there exist an open neighbourhood
$V$ of $a$ and an open neighbourhood $W$ of $b$ such that
$t(a,V)\subseteq U$ and $t(W,b)\subseteq U$.
The condition $b\in V$ or $a\in W$
contradicts the hypothesis that $t(a,b)\notin U$.

Assume now there exists a neighbourhood $U$ of $t(a,b)$ with $c\notin U$.
From $t(a,b)\in U$ and from the
separated continuity of $s$ it follows that there exist an open neighbourhood
$V$ of $a$ and an open neighbourhood $W$ of $b$ such that
$t(a,W)\subseteq U$ and $t(V,b)\subseteq U$. The condition $b\in V$ or $a\in W$
contradicts the hypothesis that $c\notin U$.
\qed

\begin{cor} Let $(\sbA,\gt)$ be an $n$-subtractive $T_0$-semitopological algebra and let $a,b\in Diag_i(\bA)$. If $s_i(a,b)\neq 0$  
then $a$ and $b$ are $T_1$-separated.
\end{cor}

\proof
 By Lemma \ref{separated} $s_i(a,b)$ and $0$ are $T_1$-separated. 
 Then the conclusion follows from Lemma \ref{semi}.
\qed

The following theorem is a slight generalization of  \cite[Theorem 5.2]{Salibra03}. 
We are indebted to \cite{Bentz99} for the technique used in the proof.

\begin{thm}\label{sep}
Let $(\sbA,\gt)$ be an $n$-subtractive $T_0$-topological algebra  and let $a,b\in Diag_i(\bA)$. 
\begin{enumerate}[label=(\roman*)]
\item If $s_i(a,b)\neq 0$, then $a$ and $b$ are $T_2$-separated in the subspace $Diag_i(\bA)$.
\item  If $s_i(a,b)$ and $0$ are
$T_2$-separated, then $a$ and $b$ are $T_{2_{1/2}}$-separated in the subspace $Diag_i(\bA)$.
\end{enumerate}
\end{thm}

\proof
(i)  By Lemma \ref{separated} there exists an open neighborhood $U$ of $s_i(a,b)$ such that $0\notin U$. From $s_i(a,b) \in U$ and from the
continuity of $s_i$ it follows that there exist an open neighborhood
$V$ of $a$ and an open neighbourhood $W$ of $b$ such that $s_i(V,W)\subseteq U$.
If there exists $d\in V\cap W\cap Diag_i(\bA)$ then $0 = s_i(d,d) \in U$ contradicting the
hypothesis.

 (ii) By hypothesis there exist
 an open neighbourhood $V'$ of $s_i(a,b)$ and an open neighbourhood $W'$ of $0$
such that $V' \cap W'= \emptyset$.
Since $s_i$ is continuous and $s_i(a,b) \in V'$, there exist two other open
sets $V$
and $W$ containing
$a$ and $b$, respectively, such that $s_i(V,W) \subseteq V'$. The pre-image of
$\overline{V'}$ under the map $s_i$
 is closed.  From  $s_i(V,W) \subseteq
V'\subseteq \overline{V'}$
the pre-image of $\overline{V'}$, that is closed, contains $V\times W$, so
$s_i(\oV,\oW) \subseteq \overline{V'}$.
We now prove that $\oV \cap \oW\cap Diag_i(\bA) = \emptyset$.
Assume, by the way of contradiction, that there is
$e\in \oV\cap\oW\cap Diag_i(\bA)$.
Since  $s_i(\oV,\oW) \subseteq \overline{V'}$ it follows that
$0 = s_i(e,e) \in \overline{V'}$. 
 But by definition of closure of a set
this is possible only if for every
 open neighbourhood $Z$ of $0$, we have that $Z\cap V' \neq \emptyset$.
But this contradicts our initial choice of $V'$ and $W'$
as two open neighbourhoods of $s(a,b)$ and $0$ respectively
with empty intersection.
\qed

\begin{cor}\label{sep2} Let $(\sbA,\gt)$ be a $2$-subtractive $T_0$-topological algebra  and let $a,b\in A$. 
 If   $s_1(a,b)\neq 0$, then  $a$ and $b$ are $T_{2_{1/2}}$-separated. In particular, for all $a\in A\setminus \{0\}$,  $a$ and $0$ are $T_{2_{1/2}}$-separated.
\end{cor}

\proof Recall that $Diag_{1}(\bA)=A$.  By Theorem \ref{sep}(i) and $s_1(s_1(a,b),0)= s_1(a,b)$ we have that $s_1(a,b)$ and $0$ are $T_2$-separated.
Then we apply Theorem \ref{sep}(ii) to get the conclusion.
\qed

We cannot generalise Theorem \ref{sep} and Corollary \ref{sep2} to semitopological algebras as explained by the following counterexample.

\begin{exa}
The Visser topology of $\gl$-calculus (see \cite{Bare, Visser80}) on the set $\gL$ of $\gl$-terms is the topology generated by the following family of sets: $U\subseteq \gL$ is a base open set if it is closed under $\gb$-conversion and it is the complement of an r.e. set. The Visser topology on the term model of the $\gl$-theory $\gl\gp\phi$  is the quotient topology of the Visser topology on $\gL$. It makes the term model of $\gl\gp\phi$  a semitopological algebra, but not a topological algebra, because Theorem \ref{sep} and Corollary \ref{sep2} are false for  the term model of $\gl\gp\phi$  with the Visser topology. In fact, the Visser topology on $\gL$ was shown hyperconnected by Visser in \cite{Visser80}. We recall that a topology is hyperconnected if the intersection of two arbitrary nonempty open sets is nonempty.
\end{exa}

We conclude the section by applying the above results to $\gl$-calculus.

\begin{cor}
 The variety of combinatory algebras generated by the term model of $\gl\gp\phi$ is of topological type  $2_{1/2}$ in $\gO$.
\end{cor}

The topological incompleteness theorem of \cite{Salibra03} states that the semantics of $\gl$-calculus given in terms of coconnected topological models   is incomplete (see Section \ref{top} for the definition of coconnected space). Coconnected topological models include all pointed ordered models of $\gl$-calculus.

In the following theorem we strongly  improve the topological incompleteness theorem of \cite{Salibra03}.
Notice that a topological model $\cM$ is not $T_{2_{1/2}}$-separated in $\gO$ if there exists an element $a$ in the model such that for all opens $U,V$ with $a\in U$ and $\gO\in V$ we have $\oU \cap \oV \neq \emptyset$.

\begin{thm}
 The semantics of $\gl$-calculus given in terms of topological models which are not $T_{2_{1/2}}$-separated in $\gO$  is theory incomplete. 
\end{thm}

%%%%%%%%%%%%%%%%%%%%%%%%%%%%%%%%%%%%%%%%%%%%%%%%%%%%%%%%%%%%%%%%%%%%%%%%%%%
\subsection{Separability in $n$-subtractive varieties}
%%%%%%%%%%%%%%%%%%%%%%%%%%%%%%%%%%%%%%%%%%%%%%%%%%%%%%%%%%%%%%%%%%%%%%%%%%%

In this last subsection we study conditions of separability for $n$-subtractive varieties of (semi)topological algebras.
 After some general results about $n$-subtractive semitopological algebras, we show how $n$-subtracivity 
 in $T_0$-topological algebras induces a generalized version of Hausdorffness due to J.P. Coleman's. Of course the focus is on
 $T_0$-topological algebras because they include the vast majority of partially ordered models of $\lambda$-calculus.
 As a particular case of this study we obtain a result stating that any $2$-subtractive $T_0$-topological algebra
 is $T_2$-separated in $0$.

\begin{defi}
  Let $\bA$ be an $n$-subtractive  algebra ($n\geq 2$) and $a\in A\setminus \{0\}$.
The \emph{rank} $\gk(a)$ of $a$ is
the least natural number $k$ such that $s_{k}(a,0) \neq 0$. 
\end{defi}

Since $s_{n-1}(a,0) = a$, then  the rank $\gk(a)$ exists
and we have $1\leq \gk(a)\leq n-1$. 

\begin{lem} $a\in Diag_{\gk(a)}(\bA)$.
\end{lem}

\proof If $\gk(a)=1$ the result follows from the identity $s_1(x,x)=0$. If $\gk(a)>1$ we have $s_{\gk(a)}(a,a) = s_{\gk(a)-1}(a,0) =0$. 
\qed

Define, for every $0\leq i \leq n-1$, $R_i = \{ a : \gk(a) \leq i\}$.
Then $R_0 =\emptyset$, $R_i\subseteq R_{i+1}$ and $R_{n-1} = A/ \{ 0\}$.

\begin{lem}\label{open1} Let $(\bA,\gt)$ be an $n$-subtractive semitopological algebra.
Then $R_i$ is open for every $i$.
\end{lem}

\proof
If $a\in A$ then $\gk(a)\leq i$ if, and only if, there is $j\leq i$ such that $s_j(a,0)\neq 0$. Then 
$$R_i = \bigcup_{j=1}^i \{a : s_j(a,0)\neq 0\}.$$
Each set $\{a : s_j(a,0)\neq 0\}$ is open because $\{0\}$ is closed, $A\setminus \{0\}$ is open and
$\{a : s_j(a,0)\neq 0\}$ is the inverse image of  $A\setminus \{0\}$ by the continuous function $s_j(x,0)$.
\qed

We define for $i\geq 1$
$$\gS_{i} = \{ a : \mbox{($\exists U,V\in \gt$)  $a\in U$, $0\in V$ and $U\cap V \subseteq R_{i-1}$} \}$$
We have: (i) $\gS_1$ is the set of all elements $a$ such that $a$ and
$0$ are $T_2$-separated; (ii) $\gS_{n} = A\setminus \{0\}$; $\gS_i \subseteq\gS_{i+1}$ for every $i$.

\begin{lem}\label{open2} Let $(\bA,\gt)$ be an $n$-subtractive semitopological algebra.
We have for every $1\leq i\leq n$:
\begin{enumerate}[label=(\roman*)]
\item $\gS_{i}$ is open.
\item $R_{i-1}\subseteq \gS_{i}$.
\end{enumerate}
\end{lem}

\proof%\hfill
%\begin{enumerate}[label=(\roman*)]
(i) Let $a\in \gS_{i}$. We have to show that there exists an open
%\item Let $a\in \gS_{i}$. We have to show that there exists an open
  neighborhood of $a$ contained within $\gS_{i}$.  If $a\in \gS_{i}$
  then the exist $U,V\in \gt$ such that $a\in U$, $0\in V$ and $U\cap
  V \subseteq R_{i-1}$. Then $U$ is an open neighborhood of $a$
  contained within $\gS_{i}$.

(ii) By Lemma \ref{open1} $R_{i-1}$ is open. Moreover, $R_{i-1}$ is
%\item By Lemma \ref{open1} $R_{i-1}$ is open. Moreover, $R_{i-1}$ is
  an open neighborhood of each of its elements and we trivially have
  $R_{i-1}\cap V \subseteq R_{i-1}$ for every (and then some) open
  neighborhood $V$ of $0$.\qed
%\end{enumerate}

\noindent In this theorem we improve Lemma \ref{open2}(ii) in the hypothesis that $\bA$ is a $T_0$-topological algebra.

\begin{thm}\label{RS} Let $(\bA,\gt)$ be an $n$-subtractive
  $T_0$-topological algebra. Then we have:
\begin{enumerate}[label=(\roman*)]
\item $a\in \gS_{\gk(a)}$ for every $a\in A\setminus \{0\}$.
\item $R_{i}\subseteq \gS_{i}$.
\item $\gS_{n-1} = A\setminus \{0\}$.
\end{enumerate}
\end{thm}

\proof %\hfill
%\begin{enumerate}[label=(\roman*)]
%\item Since  $s_{\gk(a)}(a,0) \neq 0$, then
(i) Since  $s_{\gk(a)}(a,0) \neq 0$, then
by Lemma \ref{separated} there exists an open neighbourhood $W$ of $s_{\gk(a)}(a,0)$ such that $0\notin W$. Then
we have
$$s_{\gk(a)}(a,0)\in W.$$
By the continuity of $s_{\gk(a)}$ there exist two open neighbourhoods $U$ and $V$
of $a$ and $0$ respectively such that
$$s_{\gk(a)}(U,V) \subseteq W.$$
 If $\gk(a) = 1$ and  there exists  $b\in U\cap V$, then $0= s_1(b,b) \in W$, contradicting the hypothesis on $W$. Then $V\cap U=\emptyset$; thus $a$ and $0$ are $T_2$-separated,
 and $a\in \gS_1$.
  
   If $\gk(a) > 1$, then $U\cap V\neq \emptyset$.  For every $b\in U\cap V$ we have that
$$s_{\gk(a)}(b,b) \in W,$$
that implies
$$s_{\gk(a) -1}(b,0) = s_{\gk(a)}(b,b) \neq 0.$$
This means that the rank of $b$ is less than the rank of $a$ for every $b\in U\cap V$. Then $U\cap V\subseteq R_{\gk(a) -1}$, so that $a\in \gS_{\gk(a)}$.

%\item Trivial by (i).
(ii) Trivial by (i).

%\item follows from (ii).\qed
(iii) follows from (ii).\qed

%\end{enumerate}

Let $(A,\gt)$ be a topological space. For every $a\in A$,
define by induction the following
family of subsets of $A$:
\begin{enumerate}
\item $\gG_0(a) = \emptyset$; 
\item $\gG_{i+1}(a) = \{ b : \exists\
\mbox{open $U,V$ with $a\in U$, $b\in V$ and $U\cap V \subseteq \gG_i(a)$} \}$.
\end{enumerate}
Coleman \cite{Coleman96,Coleman97} defines $(A,\gt)$ to be \emph{$n$-step Hausdorff} if   $\gG_n(a) =  A\setminus\{ a\}$ for all $a\in A$. $1$-step Hausdorff
is equivalent to $T_2$. A variety of algebras is $n$-step Hausdorff if every topological algebra in the variety is $n$-step Hausdorff. 

A variety $\cV$ of algebras is $n$-permutable ($n\geq 2$) iff every algebra in $\cV$ satisfies the Mal'cev identities of Theorem \ref{thm-unord} (see \cite{BS}).
Every $n$-permutable variety has been shown to be $\lfloor n/2\rfloor$-step Hausdorff  by Kearnes and Sequeira \cite{KearnesS02}. 

We define  a  topological algebra $(\sbA,\gt)$ to be \emph{$n$-step Hausdorff in $0$} if $\gG_n(0) = A/\{ 0\}$. $1$-step Hausdorff in $0$
is equivalent to $T_2$-separated in $0$.

\begin{prop} Let $(\sbA,\gt)$ be an $n$-subtractive $T_0$-topological algebra. Then $(\sbA,\gt)$ is $n-1$-step Hausdorff in $0$.
\end{prop}

\proof We show by induction that $\gS_i \subseteq \gG_i(0)$ for all $1\leq i \leq n$. For $i=0$ the result is trivial.
\[
\begin{array}{llll}
\gS_{i+1}  & =  & \{ b : \exists\
\mbox{open $U,V$ with $a\in U$, $b\in V$ and $U\cap V \subseteq R_i$} \}&\text{by definition}  \\
  & \subseteq  &  \{ b : \exists\
\mbox{open $U,V$ with $a\in U$, $b\in V$ and $U\cap V \subseteq \gS_i$} \} &\text{by Thm. \ref{RS}(ii)}   \\
  &  \subseteq   &   \{ b : \exists\
\mbox{open $U,V$ with $a\in U$, $b\in V$ and $U\cap V \subseteq \gG_i(0)$} \} &\text{by induction hypothesis}\\
  &  =   &  \gG_{i+1}(0) &\text{by definition}
\end{array}
\]
The conclusion follows because $\gS_{n-1}=A\setminus\{0\}$.
\qed

\end{document}

%% file: macro-salibra.tex
\newcommand{\rmA}{{\mathrm A}}
\newcommand{\rmCL}{{\mathrm{CL}}}

\newcommand{\nat}{{\mathbb{N}}}

% Macro 

\newcommand{\cA}{{\mathcal A}}
\newcommand{\cB}{{\mathcal B}}
\newcommand{\Cc}{{\mathcal C}}
\newcommand{\cD}{{\mathcal D}}
\newcommand{\cE}{{\mathcal E}}
\newcommand{\cF}{{\mathcal F}}
\newcommand{\cG}{{\mathcal G}}
\newcommand{\cH}{{\mathcal H}}
\newcommand{\cL}{{\mathcal L}}
\newcommand{\cM}{{\mathcal M}}
\newcommand{\cP}{{\mathcal P}}
\newcommand{\cO}{{\mathcal O}}
\newcommand{\cR}{{\mathcal R}}
\newcommand{\cQ}{{\mathcal Q}}
\newcommand{\cS}{{\mathcal S}}
\newcommand{\cU}{{\mathcal U}}
\newcommand{\cT}{{\mathcal T}}
\newcommand{\cV}{{\mathcal V}}

\newcommand{\la}{\langle}
\newcommand{\ra}{\rangle}

%definitions for \bf letters
\newcommand{\bb}{{\bf b }}
\newcommand{\bA}{{\bf A}}
\newcommand{\bB}{{\bf B}}
\newcommand{\bC}{{\bf C}}
\newcommand{\bD}{{\bf D}}
\newcommand{\bE}{{\bf E}}
\newcommand{\bF}{{\bf F}}
\newcommand{\bK}{{\bf K}}
\newcommand{\bM}{{\bf M}}
\newcommand{\bN}{{\bf N}}
\newcommand{\bP}{{\bf P}}
\newcommand{\bR}{{\bf R}}
\newcommand{\bS}{{\bf S}}
\newcommand{\bT}{{\bf T}}
\newcommand{\bX}{{\bf X}}
\newcommand{\bRel}{{\bf Rel}}
\newcommand{\bx}{{\bf x }}

%definitions for lower case greek letters
\newcommand{\ga}{\alpha}
\newcommand{\gb}{\beta}
\newcommand{\gam}{\gamma}
\newcommand{\gd}{\delta}
\newcommand{\gep}{\varepsilon}
\newcommand{\gz}{\zeta}
\newcommand{\geta}{\eta}
\newcommand{\gth}{\vartheta}
\newcommand{\gi}{\iota}
\newcommand{\gk}{\kappa}
\newcommand{\gl}{\lambda}
\newcommand{\gm}{\mu}
\newcommand{\gn}{\nu}
\newcommand{\gx}{\xi}
\newcommand{\gp}{\pi}
\newcommand{\gr}{\rho}
\newcommand{\gs}{\sigma}
\newcommand{\gt}{\tau}
\newcommand{\gu}{\upsilon}
\newcommand{\gph}{\varphi}
\newcommand{\gch}{\chi}
\newcommand{\gps}{\psi}
\newcommand{\go}{\omega}

\newcommand{\gG}{\Gamma}
\newcommand{\gF}{\Phi}
\newcommand{\gD}{\Delta}
\newcommand{\gT}{\Theta}
\newcommand{\gP}{\Pi}
\newcommand{\gX}{\Xi}
\newcommand{\gS}{\Sigma}
\newcommand{\gO}{\Omega}
\newcommand{\gL}{\Lambda}

\newcommand\K{\mbox{\sf K}}
\newcommand\G{\mbox{\sf G}}

\newcommand\CA{\mbox{\sf CA}}
\newcommand\LA{\mbox{\sf LA}}
\newcommand\LM{\mbox{\sf LM}}
\newcommand{\CL}{\mbox{ \sf CL}}

\newcommand\ssk{{\bf K}}
\newcommand\sss{{\bf S}}
\newcommand\ssm{\mbox{\boldmath $m$}}
\newcommand\ssi{{\bf I}}
\newcommand\ssl{\mbox{\boldmath $l$}}
\newcommand\sso{\mbox{\boldmath $1$}}
\newcommand\sst{\mbox{\boldmath $2$}}
\newcommand\ssT{\mbox{\boldmath $T$}}
\newcommand\ssF{\mbox{\boldmath $F$}}
\newcommand\ccc{\mbox{\boldmath $c$}}

\newcommand\Con{\mbox{Con}}

\newcommand\sbA{{\mathbf A}}%
\newcommand\sbB{{\mathbf B}}%
\newcommand\sbC{{\mathbf C}}%
\newcommand\sbD{{\mathbf D}}%
\newcommand\sbE{{\mathbf E}}%
\newcommand\sbF{{\mathbf F}}%
\newcommand\sbG{{\mathbf G}}%
\newcommand\sbK{{\mathbf K}}%
\newcommand\sbL{{\mathbf L}}%
\newcommand\sbM{{\mathbf  M}}%
\newcommand\sbN{{\mathbf  N}}%
\newcommand\sbO{{\mathbf O}}%
\newcommand\sbP{{\mathbf P}}%
\newcommand\sbQ{{\mathbf Q}}%
\newcommand\sbR{{\mathbf R}}%
\newcommand\sbS{{\mathbf S}}%
\newcommand\sbT{{\mathbf T}}%
\newcommand\sbU{{\mathbf U}}%
\newcommand\sbV{{\mathbf V}}%
\newcommand\sbW{{\mathbf W}}%
\newcommand\sbX{{\mathbf X}}%
\newcommand\sbY{{\mathbf Y}}%
\newcommand\sbZ{{\mathbf Z}}%

\newcommand\oa{{\overline{a}}}
\newcommand\ob{{\overline{b}}}
\newcommand\oc{{\overline{c}}}
\newcommand\od{{\overline{d}}}
\newcommand\og{{\overline{g}}}
\newcommand\ox{{\overline{x}}}
\newcommand\oy{{\overline{y}}}
\newcommand\ou{{\overline{u}}}
\newcommand\ov{{\overline{v}}}
\newcommand\ow{{\overline{w}}}
\newcommand\oz{{\overline{z}}}
\newcommand\oX{{\overline{X}}}
\newcommand\ogx{{\overline{\gx}}}
\newcommand\oA{{\overline{A}}}
\newcommand\oB{{\overline{B}}}
\newcommand\oC{{\overline{C}}}
\newcommand\oD{{\overline{D}}}
\newcommand{\oE}{{\overline E}}
\newcommand\oM{{\overline{M}}}
\newcommand\oN{{\overline{N}}}
\newcommand\oO{{\overline{O}}}
\newcommand\oP{{\overline{P}}}
\newcommand\oQ{{\overline{Q}}}
\newcommand\oU{{\overline{U}}}
\newcommand\oV{{\overline{V}}}
\newcommand\oT{{\overline{T}}}
\newcommand\oW{{\overline{W}}}

\newcommand\up{{\uparrow}}
\newcommand\down{{\downarrow}}
\newcommand\updown{{\updownarrow}}

% macro di Alberto 

\newcommand{\labeleq}[1]{=_{#1}} % equality with label
\newcommand{\labelneq}[1]{\neq_{#1}} % disequality with label
\newcommand{\labella}[1]{\ _{#1}\!\leftarrow} % left arrow with label
\newcommand{\labelra}[1]{\rightarrow_{#1}} % right arrow with label
\newcommand{\mslabella}[1]{\ _{#1}\!\twoheadleftarrow} % left two head arrow with label
\newcommand{\mslabelra}[1]{\twoheadrightarrow_{#1}} % right two head arrow with label
\newcommand{\labelmt}[1]{\mapsto_{#1}} % labelled mapsto arrow for axiomatic rewrite rules